\pdfoutput=1

\documentclass[11pt,nofootinbib]{article}
\usepackage{amsmath,amssymb,mathtools}
\usepackage{graphicx}
\usepackage[colorlinks=true,linkcolor=blue,citecolor=blue]{hyperref}
\usepackage{subfigure}
\usepackage{comment}

\numberwithin{equation}{section}

\begin{document}

\providecommand{\abs}[1]{\lvert#1\rvert}
\providecommand{\bd}[1]{\boldsymbol{#1}}

\begin{titlepage}

\setcounter{page}{1} \baselineskip=15.5pt \thispagestyle{empty}

\begin{flushright}
\end{flushright}
\vfil


\bigskip
\begin{center}
 {\LARGE \textbf{Schwinger Effect in 4D de Sitter Space and}}\\
\medskip 
 {\LARGE \textbf{Constraints on Magnetogenesis in the Early Universe}}
\vskip 15pt
\end{center}

\vspace{0.5cm}
\begin{center}
{\Large 
Takeshi Kobayashi$^{\star, \dagger}$\footnote{takeshi@cita.utoronto.ca}
and
Niayesh Afshordi$^{\dagger, \ddagger}$\footnote{nafshordi@pitp.ca}
}\end{center}

\vspace{0.3cm}

\begin{center}
\textit{$^{\star}$ Canadian Institute for Theoretical Astrophysics,
 University of Toronto, \\ 60 St. George Street, Toronto, Ontario M5S
 3H8, Canada}\\

\vskip 14pt
\textit{$^{\dagger}$ Perimeter Institute for Theoretical Physics, \\ 
 31 Caroline Street North, Waterloo, Ontario N2L 2Y5, Canada}\\

\vskip 14pt
\textit{$^{\ddagger}$ Department of Physics and Astronomy, University of
 Waterloo, \\ 200 University Avenue West, Waterloo, Ontario, N2L 3G1, Canada}

\end{center} 



\vspace{1cm}

\noindent
We investigate pair creation by an electric field in four-dimensional de
 Sitter space. The expectation value of the induced current is computed,
 using the method of adiabatic regularization. Under strong electric
 fields the behavior of the current is similar to that in flat space,
 while under weak electric fields the current becomes inversely
 proportional to the mass squared of the charged field. Thus we find
 that the de Sitter space obtains a large conductivity under weak
 electric fields in the presence of a charged field with a tiny mass.
 We then apply the results to constrain electromagnetic fields in the early
 universe. In particular, we study cosmological scenarios for
 generating large-scale magnetic fields during the inflationary
 era. Electric fields generated along with the magnetic fields 
 can induce sufficiently large conductivity to terminate the
 phase of magnetogenesis. For inflationary magnetogenesis models with a modified
 Maxwell kinetic term, the generated magnetic fields cannot exceed
 $10^{-30}\, \mathrm{G}$ on Mpc scales in the present epoch, when a
 charged field carrying an elementary charge with mass of order the
 Hubble scale or smaller exists in the Lagrangian. Similar constraints
 from the Schwinger effect apply for other magnetogenesis mechanisms.

\vfil

\end{titlepage}

\newpage
\tableofcontents

\section{Introduction}
\label{sec:intro}

Particle creation by a time dependent background happens in various
situations. The well-known example is the production of charged
particles under strong electric fields~\cite{Sauter:1931zz,Heisenberg:1935qt,Schwinger:1951nm,Nikishov:1970br,Narozhnyi:1970uv},
arising from a time dependent vector potential,
as studied by Schwinger. 
Similar phenomena are also seen in curved spacetimes, where time
dependent gravitational backgrounds produce particles. 
Such gravitational effects are particularly important for cosmology, as
the large-scale structure in the universe can be seeded by the
accelerated expansion during the inflationary epoch (see
e.g.~\cite{Martin:2007bw} for a review).
In addition to the cosmic structures, the magnetic fields in our
universe may also have a cosmological origin.
The possibility of electromagnetic fields existing in the early
universe motivates us to look into effects induced by
electromagnetic fields in curved spacetimes. 
Recently, the Schwinger effect in two-dimensional de Sitter (dS) space was
studied in~\cite{Frob:2014zka} (see also~\cite{Garriga:1994bm}).
The authors found behaviors quite different from those in flat space;
for instance, a large current is induced under weak electric fields,
when the mass of the charged particle is much smaller than the Hubble
scale, a phenomenon dubbed as ``hyperconductivity.''

With cosmological applications in mind, in this paper we explore the
Schwinger process in four-dimensional dS space. 
Considering a charged scalar, both the electric and gravitational
background fields give rise to the production of the scalar particles.
Our strategy is to study the expectation value of the induced
current, as in \cite{Frob:2014zka,Cooper:1989kf,Kluger:1991ib,Kluger:1992gb,Cooper:1992hw,Anderson:2013ila,Anderson:2013zia}.
This allows us to analyze cases where even the adiabatic vacuum does
not exist in the asymptotic future, in other words, regimes where the
scalar mass and electric force are much smaller than the Hubble scale
and thus the scalar excitations are not well described as particles.
Upon computing the expectation value of the current whose formal
expression has ultraviolet divergences, we use the method of adiabatic
subtraction~\cite{Parker:1974qw,Fulling:1974zr,Fulling:1974pu,Bunch:1978gb,Bunch:1980vc,Anderson:1987yt}
in order to remove the infinities.
Under strong electric fields, i.e. $\abs{eE} \gg H^2$, 
the induced current~$J$ is obtained as
\begin{equation}
 J \propto \frac{e^3 E^2}{H} e^{-\frac{\pi m^2}{\abs{eE}}},
\label{J1.1}
\end{equation}
where $H$ is the Hubble rate, $E$ the electric field amplitude, 
$e$ the scalar charge, and $m$ is the scalar mass.
Such a behavior of the current is analogous to that from the Schwinger
process in flat space~\cite{Anderson:2013ila,Anderson:2013zia}. 
On the other hand, with weak electric fields, i.e. $\abs{eE} \ll H^2$, 
we find that the current depends linearly on~$E$,
\begin{equation}
 J \propto \frac{e^2 E H^3}{m^2}.
\label{J1.2}
\end{equation}
Thus we confirm that for small mass, a four-dimensional dS also induces large
current from weak electric fields. 
However, unlike in the two-dimensional case~\cite{Frob:2014zka} where
the current under weak electric fields is exponentially suppressed for
massive scalars, in four-dimensions the scaling (\ref{J1.2}) holds for
arbitrary masses. 
Therefore charged massive scalars can also give rise to non-negligible
conductivity in a dS universe under weak electric fields.

After analyzing the Schwinger effect in de Sitter space, we move on to
apply the results to constrain electromagnetic fields in the
early universe. We particularly focus on cosmological scenarios for
generating large-scale magnetic fields during the inflationary epoch~\cite{Turner:1987bw,Ratra:1991bn}.
Inflationary magnetogenesis is generically accompanied by the
generation of large electric fields as
well~\cite{Bamba:2003av,Demozzi:2009fu,Kanno:2009ei,Fujita:2012rb,Kobayashi:2014sga},
which gives rise to a current via the Schwinger process.
When the induced current becomes large, its backreaction to the
Maxwell fields becomes non-negligible and can prevent any further
generation of the magnetic fields. 
Such considerations allow us to constrain models of inflationary
magnetogenesis from the Schwinger effect.
Focusing on models where the electromagnetic fields are generated by a
time dependent coupling on the Maxwell kinetic term (of the form $I(t)^2
F_{\mu \nu} F^{\mu \mu}$~\cite{Ratra:1991bn}), we find that the
Schwinger effect presents a serious obstacle to generating primordial
magnetic fields during inflation. 
For example, having a field in the action that carries an electric charge
of order the elementary charge and mass of order the Hubble scale or
smaller, the Schwinger effect prohibits inflationary magnetogenesis from 
producing magnetic fields larger than $10^{-30}\, \mathrm{G}$ on Mpc
scales in the current universe. 
The bound depends on the charges and masses of the fields in the
action, however the Schwinger effect is shown to pose a major
challenge for generating magnetic fields as large as~$10^{-15}\,
\mathrm{G}$, which is the lower bound on the extragalactic magnetic fields 
suggested by the recent gamma ray observations~\cite{Tavecchio:2010mk,Neronov:1900zz,Ando:2010rb,Taylor:2011bn,Takahashi:2013uoa,Finke:2013bua,Tashiro:2013ita}.

This paper is organized as follows:
We investigate the Schwinger effect in a four-dimensional dS
space in Section~\ref{sec:Schwinger}.
After explaining the setup, 
we carry out the usual Bogoliubov calculations in Subsection~\ref{subsec:PPR},
limiting ourselves to the
regime of $\abs{e E}, \, m^2 \gg H^2$ so that the adiabatic vacuum exists
in the asymptotic future. The reader interested in the induced
current/conductivity or constraints on magnetogenesis 
can skip this subsection, as the results obtained from the Bogoliubov
calculation will only be used upon making semiclassical estimates in
later discussions. 
In Subsection~\ref{subsec:current}, we compute the expectation value of
the current, using the method of adiabatic regularization.
The behavior of the induced current is studied in various limits,
including regimes where the Hubble scale is much larger than the 
electric force and the scalar mass.
We then apply the results to constrain inflationary magnetogenesis in
Section~\ref{sec:magnetogenesis}. This section can also be 
considered as providing discussions on the issue of backreaction to the
background electric field, in the context of magnetogenesis scenarios.
Finally, we conclude in Section~\ref{sec:conc}.

Throughout this paper, we take the principal values $-\pi
\leq \arg \varpi \leq \pi $ for the phase of complex numbers~$\varpi$.

\section{Schwinger Effect in de Sitter Space}
\label{sec:Schwinger}

In order to study the Schwinger process in a four-dimensional dS
space, we analyze QED coupled to a charged complex scalar:
\begin{equation}
 S = \int d^4 x \sqrt{-g} 
\left\{
-g^{\mu \nu} \left( \partial_\mu - i e A_\mu  \right) \varphi^* 
\left( \partial_\nu + i e A_\nu  \right) \varphi
- m^2 \varphi^* \varphi 
- \frac{1}{4} F_{\mu \nu} F^{\mu \nu}
\right\},
\label{action}
\end{equation}
where $F_{\mu \nu } = \partial_\mu A_\nu - \partial_\nu A_\mu $.
The background spacetime is fixed to dS,
\begin{equation}
 ds^2 = a(\tau)^2 \left( -d\tau^2 + dx^2 + dy^2 + dz^2 \right),
\label{FRW}
\end{equation}
where the conformal time~$\tau$ is expressed in terms of the constant
Hubble parameter as
\begin{equation}
 \tau = -\frac{1}{a H} < 0 ,
\qquad
 H = \frac{da}{a^2 d \tau} = \mathrm{const.}
\label{taudS}
\end{equation}
Here we have taken $\tau \to 0^-$ to denote the asymptotic future. 
We use Greek letters for the spacetime indices $\mu, \nu = \tau, x, y, z$,
and Latin letters for spatial indices $i, j = x, y, z$.

In order to describe a constant and uniform electric field, 
we consider a vector potential of the form
\begin{equation}
 A_\mu = 
\frac{E}{H^2 \tau } \delta_{\mu}^z,
\qquad
E = \mathrm{const.}
\label{AzE}
\end{equation}
Then a comoving observer with $4$-velocity~$u^\mu$ ($u^i = 0$, 
$u_\mu u^\mu = -1$) measures an electric field along the $z$-direction,
\begin{equation}
 E_\mu =  u^\nu F_{\mu \nu} = aE \delta_{\mu}^z,
\end{equation}
with a constant field strength $ E_\mu E^\mu = E^2$.

The equation of motion of~$\varphi$ under the time dependent
background is
\begin{equation}
 \varphi'' + 2 \frac{a'}{a}\varphi' 
- \partial_i \partial_i \varphi
-2 i e A_z \partial_z \varphi + e^2 A_z^2 \varphi + a^2 m^2 \varphi = 0,
\label{EOM1}
\end{equation}
where the prime represents a $\tau$-derivative, and the sum over
repeated spatial indices is implied irrespective of their positions.
Upon quantizing the scalar field~$\varphi$ under the time dependent
background, let us redefine the field as
\begin{equation}
 q = a \varphi  ,
\end{equation}
then the conjugate momenta are obtained from the 
action $S = \int d^4x \mathcal{L}$ in~(\ref{action}) as
\begin{equation}
 \Pi = \frac{\partial \mathcal{L}}{\partial q'} = q'^* - \frac{a'}{a}q^*,
\qquad
 \Pi^* = \frac{\partial \mathcal{L}}{\partial q'^*} = q' - \frac{a'}{a}q.
\end{equation}
We promote $q$, $q^*$, and their conjugate momenta into operators,
\begin{equation}
\begin{split}
 q(\tau, \bd{x}) & = \frac{1}{(2 \pi)^3}
\int d^3 k
\left\{  
a_{\bd{k}} q_{\bd{k}} (\tau) e^{i \bd{k}\cdot \bd{x}}
+ b_{\bd{k}}^\dagger q_{-\bd{k}}^* (\tau) e^{-i \bd{k}\cdot \bd{x}}
\right\},
\\
 q^\dagger(\tau, \bd{x}) & = \frac{1}{(2 \pi)^3}
\int d^3 k
\left\{  
a_{\bd{k}}^\dagger q_{\bd{k}}^* (\tau) e^{-i \bd{k}\cdot \bd{x}}
+ b_{\bd{k}}q_{-\bd{k}} (\tau) e^{i \bd{k}\cdot \bd{x}}
\right\},
\label{qqdagger}
\end{split}
\end{equation}
and assign the commutation relations
\begin{equation}
\begin{split}
 &[ a_{\boldsymbol{k}},\,  a_{\boldsymbol{p}}^\dagger ] =
 [ b_{\boldsymbol{k}},\,  b_{\boldsymbol{p}}^\dagger ] =
(2  \pi)^3 \, 
\delta^{(3)}  (\boldsymbol{k} - \boldsymbol{p}) , 
\\
& [ a_{\boldsymbol{k}},\,  a_{\boldsymbol{p}}] = 
 [ b_{\boldsymbol{k}},\,  b_{\boldsymbol{p}}] = 
 [ a_{\boldsymbol{k}},\,  b_{\boldsymbol{p}}] = 
 [ a_{\boldsymbol{k}},\,  b_{\boldsymbol{p}}^\dagger] = 
\cdots = 0,
\label{commu1}
\end{split}
\end{equation}
as well as
\begin{equation}
\begin{split}
& [q(\tau, \bd{x}),\, \Pi(\tau, \bd{y})] =
 [q^\dagger(\tau, \bd{x}),\, \Pi^\dagger(\tau, \bd{y})] = i
 \delta^{(3)}(\bd{x} - \bd{y}),
\\
&  [q(\tau, \bd{x}),\, q(\tau, \bd{y})] =
  [\Pi(\tau, \bd{x}),\, \Pi(\tau, \bd{y})] =
  [q(\tau, \bd{x}),\, q^\dagger(\tau, \bd{y})] =
  [q(\tau, \bd{x}),\, \Pi^\dagger(\tau, \bd{y})] =
\cdots = 0.
\label{commu2}
\end{split}
\end{equation}
The relations (\ref{commu2}) follow from (\ref{commu1}) when the 
mode function~$q_{\bd{k}}$ satisfies the normalization condition:
\begin{equation}
 q_{\bd{k}} q_{\bd{k}}'^* -  q_{\bd{k}}^* q_{\bd{k}}' =i.
\label{normcond}
\end{equation}
The mode functions obey the equation of motion (cf.~(\ref{EOM1})) taking
the form of
\begin{equation}
q_{\bd{k}}'' + \omega_{\bd{k}}^2 q_{\bd{k}} = 0,
\label{EOM2}
\end{equation}
where the effective frequency squared~$\omega_{\bd{k}}^2  $ is
\begin{align}
 \omega_{\bd{k}}^2  
 & =
(k_z + e A_z)^2 + k_x^2 + k_y^2 + a^2 m^2 - \frac{a''}{a}
\label{eq:omega_k_0}
\\
& = 
\frac{1}{\tau^2}
\left( \frac{e^2 E^2}{H^4} + \frac{m^2}{H^2} - 2 \right)
 + \frac{2}{\tau } 
\frac{k_z e E}{H^2}
+ k^2.
\label{eq:omega_k}
\end{align}
Here, $k = (k_x^2 + k_y^2 + k_z^2)^{1/2}$. 
In the asymptotic past $\tau \to - \infty$, the frequency is $
\omega_{\bd{k}}^2 \simeq k^2$, and thus $q_{\bd{k}}$ is a sum of plane waves. 
On the other hand, in the asymptotic future $\tau \to 0$, 
the frequency approaches
\begin{equation}
 \omega_{\bd{k}}^2 \simeq  \frac{1}{\tau^2} 
\left( \frac{e^2 E^2}{H^4} + \frac{m^2}{H^2} - 2 \right), 
\label{omegakastau0}
\end{equation}
whose rate of change is
\begin{equation}
 \left(\frac{\omega_{\boldsymbol{k}}'}{\omega_{\boldsymbol{k}}^2}\right)^2 
\simeq
\left( \frac{e^2 E^2}{H^4} + \frac{m^2}{H^2} - 2 \right)^{-1}, 
\qquad
\frac{\omega_{\boldsymbol{k}}''}{\omega_{\boldsymbol{k}}^3} 
\simeq
2 \left( \frac{e^2 E^2}{H^4} + \frac{m^2}{H^2} - 2 \right)^{-1}.
\label{change_rate}
\end{equation}
Thus when $e^2 E^2 / H^4 + m^2/H^2$ is much larger than unity,
then $q_{\bd{k}}$ in the asymptotic future is well
approximated by a WKB solution, in other words, there exists an adiabatic vacuum
for~$\varphi$. 

Let us now introduce the variables
\begin{equation}
 z \equiv 2 k i \tau,
\qquad
\kappa \equiv -i \frac{k_z}{k}\frac{eE}{H^2},
\qquad
\mu^2 \equiv \frac{9}{4} - \frac{e^2 E^2}{H^4} - \frac{m^2}{H^2},
\label{zkappamu}
\end{equation}
where $z$ and $\kappa$ are purely imaginary, while $\mu$ is either real
or purely imaginary. 
Then the equation of motion (\ref{EOM2}) is rewritten as
\begin{equation}
\frac{d^2 q_{\bd{k}}}{dz^2} + 
\left\{
\frac{1}{z^2}\left( \frac{1}{4} - \mu^2 \right) + \frac{\kappa }{z}
 - \frac{1}{4} 
\right\} q_{\bd{k}} = 0.
\label{eq:Whittaker}
\end{equation}
Solutions of this equation are the Whittaker functions
$  W_{\kappa, \mu} (z)$, $ M_{\kappa, \mu} (z)$, whose basic properties
are laid out in Appendix~\ref{app:Whittaker}.
From the limiting form of $W_{\kappa, \mu} (z)$
as $\abs{z} \to \infty$ shown in~(\ref{Wlimitzinf}),
we see that the function $W_{\kappa, \mu} (z)$ represents the positive 
frequency solution in the asymptotic past. Thus we choose the 
mode function as
\begin{equation}
 q_{\bd{k}} = \frac{e^{i \kappa \pi / 2}}{\sqrt{2k}} W_{\kappa, \mu}(z),
\label{qkwithW}
\end{equation}
where the normalization is set from the condition~(\ref{normcond}),
up to an arbitrary phase.

\subsection{Pair Production Rate}
\label{subsec:PPR}

Let us now evaluate the pair creation rate of the charged scalar particles. 
In this subsection we limit ourselves to cases where 
\begin{equation}
 \frac{e^2 E^2}{H^4} + \frac{m^2}{H^2} \gg 1,
\label{cond2.1}
\end{equation}
so that there exists an adiabatic vacuum for the charged
scalar in the asymptotic future.
(See discussions around~(\ref{change_rate}).)
Then the scalar excitations can be interpreted as creation of particles
at some intermediate time, and the production rate can be obtained
by computing the Bogoliubov coefficients.
Under~(\ref{cond2.1}), $\mu$ is purely imaginary,
and we take $\arg \mu = \pi / 2$, i.e.
\begin{equation}
 \mu = i \abs{\mu},
\end{equation}
throughout this subsection. 
In order to study the particle excitations at late times, let us now
rewrite the mode function in terms of~$M_{\kappa, \mu}
(z)$ (see also Appendix~\ref{app:Whittaker}, and note especially that $2 \mu$ is not an
integer in this subsection),
\begin{equation}
 q_{\bd{k}} = 
\frac{ e^{-\abs{\mu} \pi /2}}{2 \sqrt{k  \abs{\mu}}}
\left\{
\alpha_{\bd{k}}
 M_{\kappa, \mu}(z)
+
\beta_{\bd{k}} 
\left(
 M_{\kappa, \mu}(z)
\right)^*
\right\}.
\label{qkwithM}
\end{equation}
The coefficients $\alpha_{\bd{k}}$ and $\beta_{\bd{k}}$
should satisfy
\begin{equation}
 \abs{\alpha_{\bd{k}}}^2 -  \abs{\beta_{\bd{k}}}^2 = 1
\end{equation}
from the normalization condition~(\ref{normcond}).
Here we remark that,
\begin{equation}
 \frac{ e^{-\abs{\mu} \pi /2}}{2 \sqrt{k  \abs{\mu}}}
 M_{\kappa, \mu}(z)
\label{WhittakerM}
\end{equation}
represents the positive frequency solution in the asymptotic future.
From the limiting form
of~$M_{\kappa, \mu} (z)$ as $z \to 0$ shown in~(\ref{Mlimitz0}), 
it can be checked that the solution~(\ref{WhittakerM}) coincides,
up to a time independent phase, with
the WKB solution in the $\tau \to 0$ limit:
\begin{equation}
 \frac{1}{\sqrt{2 \abs{\omega_{\bd{k}}}}} 
 \exp \left\{  
- i \int^\tau  d\tau \,  \abs{\omega_{\bd{k}}} 
\right\}
\simeq
(2 \abs{\mu})^{-1/2}
(-\tau)^{i \abs{\mu} + 1/2}
\, e^{i \cdot \mathrm{const.}}.
\end{equation}
Here, upon obtaining the right hand side, we have used
(\ref{omegakastau0}) and $\abs{\mu}^2 \gg 1$. 

The Bogoliubov coefficients $\alpha_{\bd{k}}$ and $\beta_{\bd{k}}$ are
obtained from (\ref{qkwithW}) and (\ref{qkwithM}) 
by using the formula~(\ref{formulaWM}) as
\begin{equation}
 \alpha_{\bd{k}} = (2 \abs{\mu})^{1/2} 
e^{( i \kappa + \abs{\mu}) \pi /2}
\frac{\Gamma(-2 \mu)}{\Gamma (\frac{1}{2} - \mu - \kappa )},
\qquad
 \beta_{\bd{k}} = -i (2 \abs{\mu})^{1/2} 
e^{( i \kappa - \abs{\mu}) \pi /2}
\frac{\Gamma(2 \mu)}{\Gamma (\frac{1}{2} + \mu - \kappa )}.
\end{equation}
Choosing the vacuum~$| \bar{0} \rangle$ in the asymptotic future by 
$ \bar{a}_{\boldsymbol{k}}|\bar{0} \rangle =
\bar{b}_{\boldsymbol{k}}|\bar{0} \rangle = 0 $ for
$^{\forall} \boldsymbol{k}$,
where
\begin{equation}
 \bar{a}_{\bd{k}} = \alpha_{\bd{k}} a_{\bd{k}} + \beta_{\bd{k}}^*
  b_{-\bd{k}}^{\dagger}, 
\qquad
 \bar{b}_{\bd{k}} = \beta_{-\bd{k}}^* a_{-\bd{k}}^{\dagger} + \alpha_{-\bd{k}}
  b_{\bd{k}},
\end{equation}
the number of created particles 
in the vacuum~$| \bar{0} \rangle$ 
with charge~$\mp e$ and comoving wave number~$\pm \bd{k}$ 
per comoving three-volume is
\begin{equation}
 \frac{\langle \bar{0} | a_{\bd{k}}^\dagger a_{\bd{k}} | \bar{0}
  \rangle}{(2\pi)^3 \int d^3 x}
=
 \frac{\langle \bar{0} | b_{-\bd{k}}^\dagger b_{-\bd{k}} | \bar{0}
  \rangle}{(2\pi)^3 \int d^3 x}
= \frac{\abs{\beta_{\bd{k}}}^2 }{(2 \pi)^3}= 
\frac{e^{2 i \kappa \pi} + e^{- 2 \abs{\mu} \pi }}{2 (2 \pi)^3
 \sinh (2 \abs{\mu} \pi )} .
\end{equation}
Integrating this expression over all wave modes gives a divergent result:
\begin{equation}
\frac{1}{(2 \pi)^3} \int d^3k \abs{\beta_{\bd{k}}}^2 = 
\frac{1}{(2 \pi)^3 \sinh (2 \abs{\mu} \pi )}
\left\{
\frac{H^2}{ eE} \sinh \left(  \frac{2 \pi eE}{H^2} \right)
+ 2 \pi e^{- 2 \abs{\mu} \pi }
\right\}
 \int^{\infty}_0 dk\, k^2,
\label{betablowup}
\end{equation}
since it denotes the number of particle pairs produced
from the infinite past to the infinite future.
Instead of the sum over all times, we are rather interested in the
produced number of pairs per unit time.

Under the condition~(\ref{cond2.1}), the rate of change of the effective
frequency~$\omega_{\bd{k}}$~(\ref{eq:omega_k}) is tiny in both the asymptotic past and
future, and thus there exist adiabatic vacua for the charged scalar.
Here, let us estimate the time of particle creation by analyzing when the
adiabaticity is violated, i.e., when $\omega_{\bd{k}}$ changes quickly,
by studying how $\abs{\omega_{\bd{k}}' / \omega_{\bd{k}}^2}$ grows in time.
The quantity $\abs{\omega_{\bd{k}}' / \omega_{\bd{k}}^2}$ vanishes in
the asymptotic past, and it approaches the 
value~(\ref{change_rate}) in the asymptotic future. 
Depending on the parameter values, the time evolution of~$\abs{\omega_{\bd{k}}' /
\omega_{\bd{k}}^2}$ may or may not exhibit peaks in the intermediate times;
e.g. for $k_z e E < 0$, then
$\abs{\omega_{\bd{k}}' / \omega_{\bd{k}}^2}$ can just monotonically grow
in time. 
In such cases where $\abs{\omega_{\bd{k}}' / \omega_{\bd{k}}^2}$
does not exhibit peaks, we can instead focus on when $\abs{\omega_{\bd{k}}' / \omega_{\bd{k}}^2}$ comes
close to taking the asymptotic value~(\ref{change_rate}).
It can be checked that, for parameter sets that satisfy $-\mu^2 \gg 1$, 
it is around the time 
\begin{equation}
 \tau  \sim - \frac{1}{k}\left( 
\abs{\mu}^2 + \frac{1}{4} 
\right)^{1/2}
\label{tau-k-est}
\end{equation}
when the quantity~$\abs{\omega_{\bd{k}}' / \omega_{\bd{k}}^2}$ exhibits
peaks, or approaches closely to its maximum value.\footnote{The
violation of the adiabaticity can also be studied in a different frame;
by redefining the field and time as $q_{\bd{k}} = a^m \chi_{\bd{k}}$, 
$d\tau = a^{2 m} ds$, such
that the form of the equation of motion~(\ref{EOM2}) is preserved.
The detailed behavior of~$\abs{\omega_{\bd{k}}' / \omega_{\bd{k}}^2}$
(e.g., whether it peaks at a certain time, or monotonically grows)
depends on the frame, however we remark that the order-of-magnitude estimate~(\ref{tau-k-est}) 
of when $\abs{\omega_{\bd{k}}' / \omega_{\bd{k}}^2}$ approaches 
its maximum value is independent of the choice of the frame.\label{foot:frame}}

Thus we make use of the rough estimate~(\ref{tau-k-est}) and translate
the $k$-integral in~(\ref{betablowup}) into a time integral, 
\begin{equation}
\frac{1}{(2 \pi)^3} \int d^3k \abs{\beta_{\bd{k}}}^2 = 
\frac{(\abs{\mu}^2 +\frac{1}{4} )^{3/2}}{(2 \pi)^3 \sinh (2 \abs{\mu} \pi )}
\left\{
\frac{H^2}{ eE} \sinh \left(  \frac{2 \pi eE}{H^2} \right)
+ 2 \pi e^{- 2 \abs{\mu} \pi }
\right\}
 \int^0_{-\infty} d\tau\, (aH)^4.
\label{betatimeint}
\end{equation}
By looking at the produced number of pairs within~$d \tau$,
and dividing by~$a^4$,
we arrive at the pair production rate, i.e.,
the number of pairs produced per unit physical four-volume, 
\begin{equation}
\Gamma = 
\frac{H^4}{(2 \pi)^3}
\frac{(\abs{\mu}^2 +\frac{1}{4} )^{3/2}}{\sinh (2 \abs{\mu} \pi )}
\left\{
\frac{H^2}{ eE} \sinh \left(  \frac{2 \pi eE}{H^2} \right)
+ 2 \pi e^{- 2 \abs{\mu} \pi }
\right\}.
\label{pprate}
\end{equation}
Since the rate~$\Gamma$ is independent of time, 
the physical number density~$n$ of pairs at time~$\tau$ is easily
computed as
\begin{equation}
 n = \frac{1}{a(\tau)^3}\int_{-\infty}^{\tau} d\tilde{\tau} \, 
a(\tilde{\tau})^4 \Gamma 
= 
\frac{\Gamma }{3 H}.
\label{n_pair}
\end{equation}
The fact that $n$ is a constant indicates that the 
Schwinger and gravitational particle creation
balances against the dilution of the number density due to the
expansion of the universe.
One sees that, when the mass and/or the electric field are large enough
to satisfy the condition~(\ref{cond2.1}), 
the $\varphi$~population is always dominated by the particles created within a
Hubble time. 

The vacuum persistence probability can also be
computed in a similar fashion from
\begin{equation}
\left| \langle \bar{0} | 0 \rangle \right|^2 
= 
\exp \left\{
 - \frac{\int d^3 x}{(2 \pi)^3} 
\int d^3 k \, \ln \left( 1 + \abs{\beta_{\bd{k}}}^2 \right)
\right\}.
\end{equation}
Here we further assume $\frac{m^2}{H^2} \geq \frac{9}{4}$ in addition
to~(\ref{cond2.1}),
and use the formula for the dilogarithm,
\begin{equation}
 -\int^z_0 ds \, \frac{ \ln (1-s)}{s} =
  \sum_{j=1}^{\infty}\frac{z^j}{j^2},
\qquad
\mathrm{for}
\, \, 
\abs{z}\leq 1,
\end{equation}
for integrating over the angular direction~$k_z / k$.
Then, converting the $k$-integral into the time integral
using~(\ref{tau-k-est}),
one can obtain the vacuum decay rate~$\Upsilon_{\mathrm{vac}}$:
\begin{equation}
\left| \langle \bar{0} | 0 \rangle \right|^2 =
\exp\left\{
-\int d^3 x \,  d\tau \, a^4 \Upsilon_{\mathrm{vac}}
\right\},
\end{equation}
as a series of the form,
\begin{equation}
\Upsilon_{\mathrm{vac}} = 
\frac{H^4}{(2 \pi)^3}
\left(\abs{\mu}^2 +\frac{1}{4} \right)^{3/2}
\sum_{ j = 1}^{\infty}
\left\{
\frac{(-1)^{j+1}}{j^2}
\frac{2 H^2}{eE} 
e^{- 2 j \abs{\mu} \pi }
\sinh \left(  \frac{2j \pi eE}{H^2} \right)
+ \frac{2 \pi }{j}
e^{- 4 j \abs{\mu} \pi }
\right\}.
\end{equation}

Let us close this section by studying the limit where the Hubble
parameter is much smaller than the mass and electric field strength.
Taking $H \to 0 $ in the above expressions gives
\begin{equation}
 \lim_{H \to 0} \Gamma  =
\frac{(eE)^2}{(2 \pi)^3} 
\exp \left( -\frac{\pi m^2}{\abs{eE}} \right),
\label{Gamma_in_Mink}
\end{equation}
\begin{equation}
 \lim_{H \to 0}    \Upsilon_{\mathrm{vac}} =
\sum_{j=1}^{\infty} 
\frac{(-1)^{j+1}}{j^2}
\frac{(eE)^2}{(2 \pi)^3}
\exp \left( - \frac{j \pi  m^2}{\abs{eE}} \right),
\end{equation}
reproducing the familiar results for Schwinger pair production in
Minkowski space~\cite{Schwinger:1951nm,Nikishov:1970br,Narozhnyi:1970uv} 
(see also~\cite{Anderson:2013ila} for a recent analysis).

\subsection{Induced Current and Conductivity}
\label{subsec:current}

Once produced, the charged scalar particles move under the electric
field and thus give rise to a current as well as a conductivity. 
The results obtained in Subsection~\ref{subsec:PPR} can be used to estimate
the induced current via $\abs{J} \sim \abs{2 e n v}$, where $v$ is the velocity of the
particles. Such a semiclassical approach provides good approximations in
some parameter regions (as we will see later), but not in general. 
In particular, computing the contribution only from the created
particles is not enough, as such a naive picture violates local charge
conservation~\cite{Frob:2014zka}.
Moreover, the analyses in the previous subsection were limited to 
cases where the mass and/or the electric force are sufficiently
larger than the Hubble scale, cf.~(\ref{cond2.1}). 

In this subsection we do {\it not} impose the condition~(\ref{cond2.1}), and 
directly compute the expectation value of the conserved
current,\footnote{In terms of the current $J_\mu$~(\ref{currentJ}),
the Maxwell equation is written as 
\begin{equation}
 \nabla^{\nu} F_{\mu \nu} = J_{\mu}.
\label{Maxwell}
\end{equation}
}
\begin{equation}
 J_\mu = \frac{i e}{2}   
\left\{
\varphi^\dagger (\partial_\mu + i e A_\mu) \varphi
 - \varphi (\partial_\mu - i e A_\mu) \varphi^\dagger
\right\}
+ \mathrm{h.c.}
\label{currentJ}
\end{equation}
in the vacuum state~$| 0 \rangle$ defined by 
$ a_{\boldsymbol{k}}| 0 \rangle = b_{\boldsymbol{k}}|0 \rangle = 0 $ for
$^{\forall} \boldsymbol{k}$, cf.~(\ref{qqdagger}).
Under the electric field along the $z$-direction, 
the expectation value vanishes except for its $z$-component,
\begin{equation}
 \langle J_z \rangle
= - \frac{2 e }{(2 \pi)^3 a^2}\int d^3 k \, 
(k_z + e A_z) \abs{q_{\bd{k}}}^2,
\label{jz2.42}
\end{equation}
where the mode function is given in~(\ref{qkwithW}).
However this expectation value diverges, as can be seen from the limiting form
of~$q_{\bd{k}}$ as $k \to \infty$ shown in~(\ref{Wlimitzinf}).
In order to explicitly see the divergence, let us first compute the integral 
by imposing a cutoff~$\zeta $ on~$k$, 
\begin{equation}
 \langle J_z \rangle =
-\lim_{\zeta \to \infty}  \frac{2 e }{(2 \pi)^2 a^2}
 \int^{\zeta}_0 dk\, k^2 
\int^1_{-1} d r\, 
 (k r  + e A_z) 
\frac{e^{i \kappa \pi }}{2 k}
\left|
W_{\kappa, \mu} (z)
\right|^2,
\label{jzWkmu}
\end{equation}
where we have introduced 
\begin{equation}
 r = \frac{k_z}{k}.
\end{equation}
The integral is carried out in Appendix~\ref{app:int}, yielding
\begin{equation}
\begin{split}
 \langle J_z \rangle =
& \frac{e a H^3 }{(2 \pi)^2}
\lim_{\zeta \to \infty} \Biggl[
\frac{2 \lambda }{3} \left(\frac{\zeta }{a H}\right)^2 + \frac{\lambda }{3} \ln
 \left(\frac{2 \zeta}{aH} \right) 
  - \frac{25\lambda }{36} + \frac{\mu^2 \lambda }{3} + \frac{\lambda^3}{15} 
\\
& +\frac{45 + 4 \pi^2 (-2+3 \lambda^2 + 2 \mu^2)}{12 \pi^3}
\frac{\mu \cosh (2 \pi \lambda )}{\lambda \sin (2 \pi \mu)}
 -\frac{45 + 8 \pi^2 (-1+9 \lambda^2 +  \mu^2)}{24 \pi^4}
\frac{\mu \sinh (2 \pi \lambda )}{\lambda^2 \sin (2 \pi \mu)}
\\
&+ \mathrm{Re}
\biggl\{
 \int^1_{-1} dr\, 
\frac{i \lambda }{16 \sin (2 \pi \mu )}
\left(
-1+4 \mu^2 + (7 + 12 \lambda^2 - 12 \mu^2) r^2 - 20 \lambda^2 r^4
\right)
\\
&\qquad \, \, \, \,  \times
\left(
\left( e^{2 \pi r \lambda }+ e^{2 \pi i \mu} \right)
\psi \left( \tfrac{1}{2} + \mu + i r \lambda  \right)
-
\left( e^{2 \pi r \lambda }+ e^{-2 \pi i \mu} \right)
\psi \left( \tfrac{1}{2} - \mu + i r \lambda  \right)
\right)
\biggr\}
\Biggr],
\label{rawJ}
\end{split}
\end{equation}
where $\psi (z) = \Gamma'(z) / \Gamma (z)$ is the digamma function, and
$\lambda$ is defined as 
\begin{equation}
 \lambda = \frac{e E}{H^2}.
\end{equation}
We thus see that the expectation value of the current has quadratic and
logarithmic divergences.  
Let us also remark that some of the terms in (\ref{rawJ}) blow
up when $\mu = 0, 1/2, \dotsc$. However their sum does not
necessarily diverge as $\mu$ approaches such values, and thus
the finite part of (\ref{rawJ}) (i.e. terms without~$\zeta$) is well-behaved.

In order to regularize the divergences, we use the method of adiabatic
subtraction~\cite{Parker:1974qw,Fulling:1974zr,Fulling:1974pu,Bunch:1978gb,Bunch:1980vc,Anderson:1987yt}.
The idea here is to compute quantities in the limit of slow variation of
the background,
then subtract their contributions from the formal expressions
to obtain a finite result.
(See also
works~\cite{Cooper:1989kf,Kluger:1991ib,Kluger:1992gb,Cooper:1992hw}
which applied adiabatic regularization to the analysis of Schwinger
effect in flat space.)
Let us start by considering a mode function with a WKB form,
\begin{equation}
 q_{\bd{k}}(\tau) = \frac{1}{\sqrt{2 W_{\bd{k}}(\tau)}}
\exp \left\{
-i \int^\tau d\tilde{\tau}\, W_{\bd{k}} (\tilde{\tau})
 \right\},
\label{WKBsolution}
\end{equation}
which is an exact solution of the equation of motion~(\ref{EOM2}) if the
function~$W_{\bd{k}}$ satisfies
\begin{equation}
 W_{\bd{k}}^2 = \omega_{\bd{k}}^2 + \frac{3}{4} \left(
 \frac{W_{\bd{k}}'}{W_{\bd{k}}}
 \right)^2
 - \frac{1}{2} \frac{W_{\bd{k}}''}{W_{\bd{k}}}.
\label{Wk-omegak}
\end{equation}
Furthermore, when $W_{\bd{k}}$ is real and positive,
the normalization condition~(\ref{normcond}) is also satisfied. 
Here, recall from (\ref{eq:omega_k_0}) that $\omega_{\bd{k}}^2$ takes
the form of
\begin{equation}
 \omega_{\bd{k}}^2 = \Omega_{\bd{k}}^2 - \frac{a''}{a},
\label{omegak-Omegak}
\end{equation}
with
\begin{equation}
 \Omega_{\bd{k}} = \left\{ (k_z + e A_z)^2 + k_x^2 + k_y^2 + a^2 m^2
		   \right\}^{1/2}.
\label{Omega_k251}
\end{equation}
Hereafter, let us assume the mass to be nonzero, i.e. $m \neq 0$, so that 
$\Omega_{\bd{k}}^2$ is positive definite.
In order to parameterize the
slowness of the evolution of the time dependent background,
we assign an adiabatic parameter~$T^{-1}$ to each time derivative in 
(\ref{Wk-omegak}) and (\ref{omegak-Omegak});
taking $T \to \infty $ denotes the 
limit of infinitely slow variation of the background.
Then the function~$W_{\bd{k}}^2$ can be computed at each order
in~$T^{-1}$.
The solution at leading order is simply
\begin{equation}
 W_{\bd{k}}^2 = \Omega_{\bd{k}}^2 + \mathcal{O}(T^{-2}). 
\label{Wkadi1}
\end{equation}
Higher order solutions can be obtained by recursively substituting
the results into the right hand side of~(\ref{Wk-omegak});
up to adiabatic order~$T^{-2}$ we obtain
\begin{equation}
 W_{\bd{k}}^2 = \omega_{k}^2 + \frac{3}{4} 
\left( \frac{\Omega_{\bd{k}}'}{\Omega_{\bd{k}}} \right)^2
- \frac{1}{2} \frac{\Omega_{\bd{k}}''}{\Omega_{\bd{k}}} 
+ \mathcal{O}(T^{-4}),
\label{Wkadi2}
\end{equation}
and so on. 
We note that our results are not altered by 
computing the adiabatic subtraction terms in a different frame, 
where $q_{\bd{k}}$ and $\tau$ are redefined
such that the equation of motion preserves the form of~(\ref{EOM2})
(cf.~Footnote~\ref{foot:frame}).

In the following we expand the current in terms of~$T^{-1}$,
then the lower order results will be subtracted from the formal expression 
(\ref{jzWkmu}).
We will see that the adiabatic subtraction up to quadratic order is just
enough to remove the divergences, as well as gives results that have the
correct behavior in the Minkowski limit.
We also remark that, since the formal expression of $\langle J_0
\rangle$ vanishes, and also $\langle J_i \rangle$ is homogeneous, it
is clear that the adiabatic subtraction does not spoil the current conservation.
Detailed discussions on the method of adiabatic subtraction can be found in, 
e.g.,~\cite{Parker:1974qw,Fulling:1974zr,Fulling:1974pu,Bunch:1978gb,Bunch:1980vc,Anderson:1987yt}. 

For a real and positive~$W_{\bd{k}}$ (recall from (\ref{Wkadi1}) that
we can have $W_{\bd{k}} \simeq \Omega_{\bd{k}} > 0$ at the leading order),
the current~(\ref{jz2.42}) is written in terms of
the WKB solution~(\ref{WKBsolution}) as
\begin{equation}
- \frac{2 e }{(2 \pi)^3 a^2} \int d^3 k\, (k_z + e A_z) 
\frac{1}{2 W_{\bd{k}}}.
\end{equation}
Expanding this expression up to adiabatic order~$T^{-2}$ using
(\ref{Wkadi2}) yields
\begin{equation}
\begin{split}
- \frac{2 e }{(2 \pi)^3 a^2}
 \int d^3 k \, (k_z + e A_z) \frac{1}{2 \Omega_{\bd{k}}}
\Biggl[
1 
&+ \frac{1}{2 \Omega_{\bd{k}}^2} \frac{a''}{a}
\\
& \quad
+ \frac{1}{4 \Omega_{\bd{k}}^4}
\left\{
(e A_z')^2 + (k_z + e A_z ) e A_z'' + 
 \left( a'^2 + a a''  \right) m^2
\right\}
\\
& \qquad \qquad 
 -\frac{5}{8 \Omega_{\bd{k}}^6}
\left\{  (  k_z + e A_z) e A_z' + a a' m^2 \right\}^2
+ \mathcal{O} (T^{-4})
\Biggr].
\label{adisub2.54}
\end{split}
\end{equation}
We carry out the integration 
by imposing a cutoff~$\zeta$ on~$k$ as in~(\ref{jzWkmu}).
After some algebra we obtain
\begin{equation}
\begin{split}
\lim_{\zeta \to \infty}
 \frac{ e }{(2 \pi)^2 a^2}
\Biggl[
&- \frac{2  }{3} e A_z \zeta^2 
+ \frac{2  }{15}(eA_z)^3 + \frac{1 }{3} e A_z a^2 m^2 
\\
&
- \frac{1 }{6} e A_z'' \ln \left( \frac{2 \zeta }{am} \right) 
+ \frac{1 }{6}\frac{a' }{a}  eA_z'
- \frac{1 }{3} \frac{a''}{a} e A_z
+ \frac{2  }{9} e A_z'' 
+ \mathcal{O}(T^{-4})
\Biggr].
\end{split}
\end{equation}
Terms of adiabatic order~$T^0$ shown in the first line contains a
quadratic divergence, while the terms of~$T^{-2}$ in the second line has
a logarithmic divergence. 
Substituting the expressions for $a$~(\ref{taudS}) and $A_z$~(\ref{AzE}),
we find the adiabatic subtraction terms to be 
\begin{equation}
\lim_{\zeta \to \infty}
\frac{ e a H^3 }{ (2 \pi)^2}
\Biggl[
\frac{2\lambda  }{3}  \left( \frac{\zeta }{a H} \right)^2
 - \frac{2\lambda^3}{15}  
- \frac{\lambda }{3} \frac{ m^2}{ H^2} 
 +\frac{\lambda }{3}
\ln \left(\frac{2  \zeta }{a m} \right) + \frac{\lambda }{18} + \mathcal{O} (T^{-4})
\Biggr],
\label{adi-sub}
\end{equation}
where the first three terms 
arise from the 
order~$T^0$ expansion, and the other two terms from order~$T^{-2}$.
Comparing with the formal expression~(\ref{rawJ}),
the divergences of the expectation value 
are seen to be removed by the adiabatic subtraction up to order~$T^{-2}$.
One can further expand up to order~$T^{-4}$, which only gives finite terms.
However, as we will see later, subtracting terms of
order~$T^{-4}$ spoils the behavior of the current in the flat space
limit. Therefore, we subtract off terms up to adiabatic order~$T^{-2}$
from~(\ref{rawJ}) in order to obtain the regularized current, arriving at
\begin{equation}
\begin{split}
\! 
 \langle J_z  \rangle_{\mathrm{reg}} 
=
 &
 \frac{e a H^3  }{(2 \pi)^2}
\Biggl[
- \frac{2 \lambda^3}{15} 
+  \frac{\lambda}{3} \ln  \left(\frac{m}{H} \right) 
\\
& 
+\frac{45 + 4 \pi^2 (-2+3 \lambda^2 + 2 \mu^2)}{12 \pi^3}
\frac{\mu \cosh (2 \pi \lambda )}{\lambda \sin (2 \pi \mu)}
 -\frac{45 + 8 \pi^2 (-1+9 \lambda^2 +  \mu^2)}{24 \pi^4}
\frac{\mu \sinh (2 \pi \lambda )}{\lambda^2 \sin (2 \pi \mu)}
\\
& 
+ \mathrm{Re}
\biggl\{
 \int^1_{-1} dr\, 
\frac{i \lambda}{16 \sin (2 \pi \mu )}
\left(
-1+4 \mu^2 + (7 + 12 \lambda^2 - 12 \mu^2) r^2 - 20 \lambda^2 r^4
\right)
\\
&\qquad \, \, \, \,  \times
\left(
\left( e^{2 \pi r \lambda }+ e^{2 \pi i \mu} \right)
\psi \left( \tfrac{1}{2} + \mu + i r \lambda  \right)
-
\left( e^{2 \pi r \lambda }+ e^{-2 \pi i \mu} \right)
\psi \left( \tfrac{1}{2} - \mu + i r \lambda  \right)
\right)
\biggr\}
\Biggr].
\label{Jreg}
\end{split}
\end{equation}
Comparing with the formal expression~(\ref{rawJ}), the procedure of
adiabatic subtraction has modified the terms in the first line inside the parentheses.

Here it is important to note that the hard cutoff~$\zeta$ was introduced
in the derivation
only for calculational convenience, so that the integrations of the formal
expression~(\ref{jzWkmu}) and the adiabatic expansion~(\ref{adisub2.54})
can be performed separately. 
Instead of using~$\zeta$, we could have subtracted the integrand of 
(\ref{adisub2.54}) from (\ref{jzWkmu}) before carrying out the integral,
then we would not see any infinities in the calculations. Such a
procedure would be preferable for numerical studies.

Before discussing the behavior of~(\ref{Jreg}), let us 
parameterize the amplitude of the current as
\begin{equation}
\langle J_z \rangle_{\mathrm{reg}}  = a J,
\label{J-def}
\end{equation}
where $J$ has mass dimension three.
We also define the conductivity~$\sigma$ by
\begin{equation}
 \sigma = \frac{J}{E}.
\label{sigma-def}
\end{equation}
Then one sees from (\ref{Jreg}) that normalized quantities 
such as 
\begin{equation}
 \frac{J}{e H^3}, \qquad 
\frac{\sigma }{e^2 H} = 
\frac{J}{e H^3 }  \frac{1}{\lambda } 
\end{equation}
are uniquely fixed by the two parameters
$m / H$ and $e E / H^2$ (or, equivalently, $\mu$ and $\lambda$),
representing the mass and electric force relative to the Hubble scale.
In particular, $J / e H^3$ and $\sigma / e^2 H$ are independent of time.
Note also that $J / e H^3$ and $\sigma / e^2 H$ are, respectively, odd and
even under $\lambda  \to - \lambda $.

\begin{figure}[htbp]
\centering
\subfigure[current]{%
  \includegraphics[width=.58\linewidth]{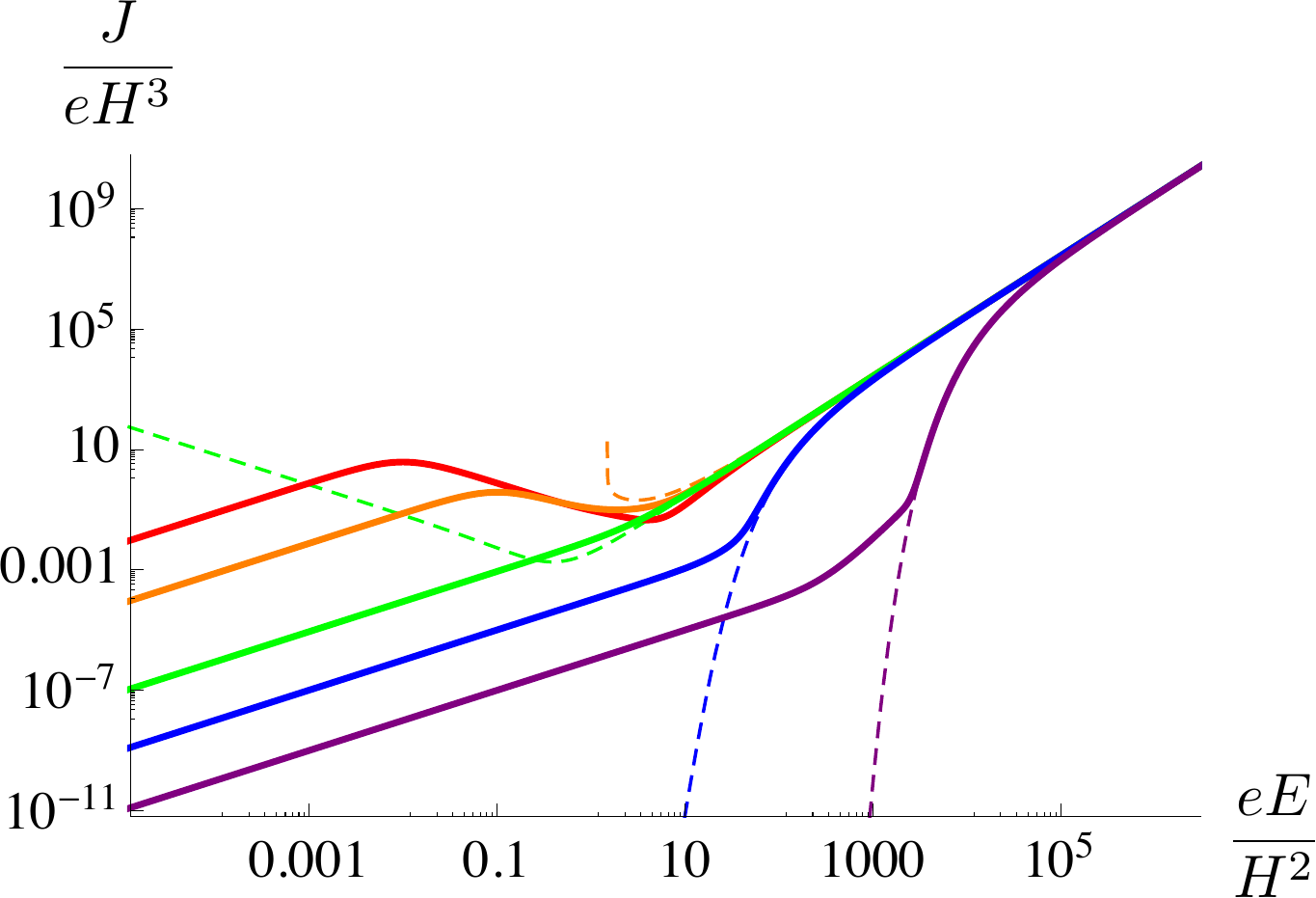}
  \label{fig-J}}
\subfigure[conductivity]{%
  \includegraphics[width=.58\linewidth]{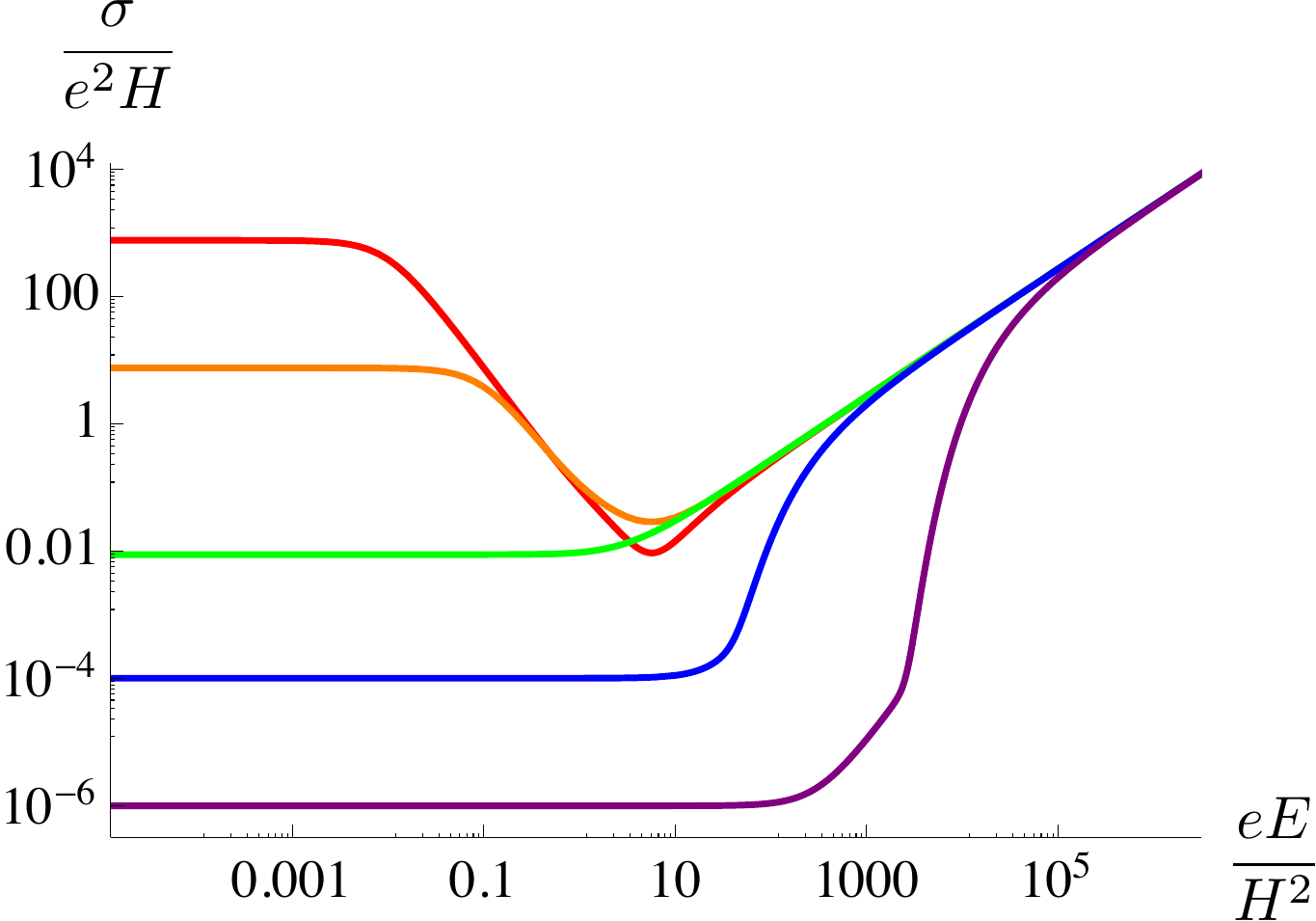}
  \label{fig-sigma}}
\caption{Induced current~$J$ and conductivity~$\sigma$  as a function of the
 electric field~$E$. The displayed quantities are normalized by the
 Hubble parameter~$H$ and charge~$e$. Each line is for a different
 choice of mass, $m / H = $ $0.01$(red), $0.1$(orange), $1.5$(green),
 $10$(blue), $100$(purple). The semiclassical estimate of the current using $\abs{J}
 \sim  2 \abs{e} n$ is shown as dashed lines, with colors representing the
 choice of mass.} 
\label{fig:J-sigma}
\end{figure}

In Figure~\ref{fig:J-sigma} we plot $J / e H^3$ and $\sigma / e^2 H$ as
functions of~$\lambda$, where each curve corresponds to a different
choice of mass~$m / H$. 
The solid lines are obtained from the regularized result~(\ref{Jreg}). 
We also show dashed lines denoting semiclassical estimates of the
current based on the computations in Subsection~\ref{subsec:PPR}, which
will be explained later.

The plots show that when $\abs{e E} \gg H^2$, the conductivity monotonically grows with
increasing~$\abs{E}$, and becomes independent of the mass 
for a sufficiently large $\abs{e E} / H^2$. 
On the other hand, the conductivity is independent of~$E$ under weak
electric force $ \abs{e E }\ll H^2$.
In particular, for small mass~$m \ll H$, the conductivity is strongly
enhanced in the weak electric field regime. 
Let us now study the behavior of $J$ and $\sigma$ in the limiting
regimes of $\abs{eE} \gg H^2$ and $\abs{eE} \ll H^2$, respectively.

\subsubsection{Strong Electric Force : $ \abs{e E}  \gg H^2$}
\label{subsec:StrongE}

In the limit of $\abs{\lambda} \to \infty$, 
for a fixed $m / H$, 
the third and fourth lines of~(\ref{Jreg}) approach
\begin{equation}
 \mathrm{Re} \left\{ \int dr \cdots \right\} \simeq \frac{2\lambda^3}{15},
\end{equation}
and thus largely cancels with the first term
in~(\ref{Jreg}). Consequently, the $\cosh$ term proportional
to~$\lambda^2$ (note that $\mu^2 \simeq - \lambda^2$)
dominates the current and yields
\begin{equation}
\frac{J}{H^3} \simeq 
\mathrm{sgn}( \lambda ) \,
\frac{e \lambda^2  }{12 \pi^3} 
= 
\mathrm{sgn}( E ) \, 
\frac{1 }{12 \pi^3}  \frac{\abs{e}^3 E^2}{H^4},
\label{largeG_J}
\end{equation}
\begin{equation}
\frac{\sigma }{H} \simeq 
\frac{e^2 \abs{\lambda } }{12 \pi^3}
= \frac{1}{12 \pi^3} \frac{\abs{e^3 E}}{H^2}.
\label{largeG_sigma}
\end{equation}
Thus one sees that when $\abs{eE}$ is sufficiently large relative to the 
Hubble rate~$H^2$, the current~$J$ is quadratic in~$E$, and thus
$\sigma$ is linear in~$E$. 
These features are also seen in the plots,
where all curves converge as $\abs{\lambda} \to \infty$ and the behavior
becomes independent of the mass.

\vspace{\baselineskip}

Since the condition (\ref{cond2.1}) is satisfied in the regime of
$\abs{\lambda} \gg 1$, an adiabatic vacuum exists in the asymptotic future
for the $\varphi$ fields.
Hence let us try to estimate the current in this regime based on the
semiclassical picture of the created particles carrying the charge.
Supposing the particles to travel with velocity $v \simeq 1$, then
we can estimate the arising current by
\begin{equation}
 J \simeq \mathrm{sgn} ( E ) \, 2 \abs{e} n = 
\mathrm{sgn} (  E  ) \, \frac{2 \abs{e} \Gamma }{3 H}.
\label{J2en}
\end{equation}
Here $n$ is the number density of the produced pairs, and we have used 
(\ref{n_pair}) upon moving to the right hand side.
The pair production rate~$\Gamma$ was computed in~(\ref{pprate}). 
Taking the limit of $H \to 0$ as shown in~(\ref{Gamma_in_Mink}), and
further taking $m^2 / eE \to 0$, one exactly reproduces the
result~(\ref{largeG_J}). 

In Figure~\ref{fig-J} the dashed lines show the semiclassical
estimate~(\ref{J2en}) 
using the production rate~$\Gamma $~(\ref{pprate}). 
In the regime $\abs{\lambda} \gg 1$, the estimates agree
well with the results obtained from 
computing~$\langle J_z \rangle_{\mathrm{reg}}$. 
For cases with $ m \gg H$, the suppression of the
current at $ \abs{eE} \sim m^2$ corresponds to the 
mass suppression of~$\Gamma$ shown in~(\ref{Gamma_in_Mink}).
Thus we obtain an improved approximation for the current
(\ref{Jreg}) at $\abs{\lambda} \gg 1$ as
\begin{equation}
 J \simeq 
\mathrm{sgn}( E)\, 
\frac{1}{12 \pi^3}  \frac{\abs{e}^3 E^2}{H}
e^{ -\frac{\pi m^2}{\abs{eE}}}.
\label{expMsupp}
\end{equation}
On the other hand, when $\abs{\lambda} \ll 1$, 
the semiclassical estimate is seen to break down. Particularly, for
light masses $ m / H < 3/2 $, the condition~(\ref{cond2.1}) does not
hold and thus the expression~(\ref{pprate}) itself cannot be
extended to the $\abs{\lambda} \ll 1$ regime. 
(We note that the semiclassical estimate for $m/H = 0.01 $ (red dashed
line) overlaps with that of $m / H = 0.1$ (orange dashed), and thus not
seen in the plot.)

\vspace{\baselineskip}

The behavior of the current~(\ref{expMsupp}) in the regime $\abs{\lambda }\gg
1$ corresponds to that induced by the Schwinger effect in a Minkowski 
space~\cite{Anderson:2013ila,Anderson:2013zia},
\begin{equation}
 J \sim \mathrm{sgn}( E)\, \abs{e}^3 E^2 (t-t_0) e^{-\frac{\pi m^2}{\abs{eE}}},
\label{Jflat}
\end{equation}
where $t_0$ is the initial time when the electric field is switched on.
(The situation here is slightly different from that in (\ref{expMsupp})
where the electric field always exist.
A finite $t_0$ is introduced because, due to the absence of the Hubble
dilution, $J$ blows up in a Minkowski space if the electric field existed from the
infinite past $t_0 = -\infty$. This is why the
expression~(\ref{expMsupp}) diverges as $H\to 0$.)
Our results in dS were obtained by an adiabatic expansion
up to quadratic order,
however we remark that the adiabatic subtraction at order~$T^{-4}$
produces a term that scales as~$\lambda^3$, which
gives a scaling behavior of $J \propto E^3$ at $\abs{\lambda} \gg 1$,
contrary to~(\ref{Jflat}).
Therefore we see that the adiabatic subtraction up to order~$T^{-2}$ not just removes
the divergences, 
but also produces results with the correct behavior in
the Minkowski limit.

Let us also remark that the scaling of~$J$ at $\abs{\lambda} \gg 1$
depends on the spacetime dimension.
In a two-dimensional dS space, the current induced by strong electric
fields scales as $J \propto E$, and thus the conductivity approaches a
constant at large~$\abs{E}$~\cite{Frob:2014zka}.
As we will see in Section~\ref{sec:magnetogenesis}, $\sigma$ being an
increasing function of~$E$ in four-dimensions gives rise to stringent
constraints on electromagnetic fields in the early universe.

\subsubsection{Weak Electric Force : $\abs{eE} \ll H^2$}
\label{subsec:WeakE}

Taking $\lambda \to 0$ for a fixed $ m / H$, one can check that the
current (\ref{Jreg}) becomes dominated by terms linear in~$\lambda$, 
and approximated as
\begin{equation}
\frac{J}{H^3 } 
\simeq
 \frac{e^2 E }{24 \pi^2 H^2}
\left\{
 \ln \left( \frac{m^2}{H^2} \right)
+ \frac{16 \pi }{3} \frac{\mu_0 (-1 + \mu_0^2)}{\sin (2 \pi
\mu_0)}
- \psi \left(\tfrac{1}{2} + \mu_0\right)
- \psi \left(\tfrac{1}{2} - \mu_0\right)
\right\},
\end{equation}
where
\begin{equation}
\mu_0^2 = \frac{9}{4} - \frac{m^2}{H^2}.
\end{equation}
The expression gets further simplified when $ m \gg H$ as
\begin{equation}
 \frac{J}{H^3} \simeq
\frac{7}{72 \pi^2}
\frac{e^2 E}{m^2}.
\end{equation}
Here the main contributions to the current are given by the
terms $\frac{\lambda }{3} \ln (\frac{m}{H})$ and 
$\mathrm{Re}\{\int dr \cdots \}$ in~(\ref{Jreg}),
which originate from the adiabatic subtraction~(\ref{adi-sub}) and
formal expression~(\ref{rawJ}), respectively. 
On the other hand, for $ m \ll H$, the current is approximated by
\begin{equation}
 \frac{J}{H^3} \simeq 
\frac{3}{4 \pi^2 }
\frac{e^2 E }{ m^2} .
\end{equation}
The current in this case mainly arises from the terms in~(\ref{Jreg})
that involve $\cosh$, $\sinh$, and $\mathrm{Re}\{\int dr \cdots \}$,
which are all contributions originating from the formal expression~(\ref{rawJ}).

Therefore for arbitrary mass, 
the current and conductivity in the regime $\abs{\lambda} \ll 1$
can be roughly approximated by 
\begin{equation}
 \frac{J}{H^3} \sim 10^{-2} \times \frac{e^2 E}{m^2},
\qquad
 \frac{\sigma }{H} \sim 10^{-2} \times \frac{e^2 H^2}{m^2}.
\label{sigma_weakE}
\end{equation}
Here we see that the conductivity is independent of~$E$, and
takes larger values for a smaller mass ratio~$ m / H$.

In particular for light mass $ m \ll H$, 
the plot shows that as one decreases~$\lambda$, 
the conductivity grows 
until it approaches the constant value~(\ref{sigma_weakE}).
The strong enhancement of the conductivity for small mass scalars
under weak electric fields are supported by the
infrared contributions to the current:
This can be seen from the limiting behavior of the Whittaker function as $z \to 0$,
\begin{equation}
 W_{\kappa, \mu } (z ) \propto z^{\frac{1}{2} - \abs{\mathrm{Re} (\mu
  )}}.
\label{Wscalingas0}
\end{equation}
(Here we note that this expression (\ref{Wscalingas0}) is not
valid for an arbitrary set of $\kappa$ 
and $\mu$, however we use it for the rough estimation in the following.)
Hence one sees that the integrand of (\ref{jzWkmu}) for $\langle J_z \rangle$,
in the limit $k \to 0$, scales as
\begin{equation}
 \propto  k^{2 ( 1 - \abs{\mathrm{Re}(\mu )})}.
\label{IRspectrum}
\end{equation}
The simple power counting estimate indicates that the current spectrum
diverges in the infrared limit when $\abs{\mathrm{Re} (\mu)} > 1$, i.e.,
\begin{equation}
 \frac{e^2 E^2}{H^4} + \frac{m^2}{H^2} < \frac{5}{4}.
\end{equation}
However, it should also be noted that a nonzero~$e E$ or $m$ give
$\abs{\mathrm{Re} (\mu)} < \frac{3}{2}$, 
and thus integrating the spectrum~(\ref{IRspectrum}) down to $k = 0$
yields a finite total current. We also remark that the spectrum of the
adiabatic subtraction terms in~(\ref{adisub2.54}) are finite in the
infrared limit, since $\Omega_{\bd{k}} \geq  a m > 0 $ for a nonzero
mass.
Thus we find that the enhanced conductivity at $\abs{\lambda} \ll 1$ for
small masses arise from the infrared behavior of the mode
function~$q_{\bd{k}}$.
The nature of the infrared modes being important should be
contrasted to cases $-\mu^2 \gg 1$ discussed below~(\ref{n_pair}), 
where the newly created particles with $k \sim a H \abs{\mu}$
always dominate the $\varphi$~population.

\vspace{\baselineskip}

The constancy of~$\sigma$ under weak electric fields, and its strong
enhancement for small mass are also seen in a two-dimensional dS
space~\cite{Frob:2014zka}. However there is an important difference worth
noting: While in four dimensions the conductivity under weak fields
scales as $\sigma \propto 
m^{-2}$ for all masses, in a two-dimensional dS the scaling $\sigma
\propto m^{-2}$ is only for light masses, and the conductivity is
exponentially suppressed for $ m \gtrsim H$.

\subsubsection{Comments on Very Light or Massless Scalars}

In Figure~\ref{fig:J-sigma} we have plotted curves for scalar masses as
low as $m / H = 0.01$.
Cases for even smaller masses have similar behaviors under strong/weak electric fields;
the conductivity takes the mass independent value~(\ref{largeG_sigma})
at $\abs{\lambda} \gg 1$, while at $\abs{\lambda} \ll 1$ the conductivity grows 
with decreasing~$\lambda$ at a more or less similar rate, until it
approaches the constant value~(\ref{sigma_weakE}) set by the mass. 
However, we remark that the valley of~$\sigma$ 
in the intermediate regime of $\lambda $ becomes deeper for smaller~$m$,
and the conductivity can even take negative values at $\abs{\lambda} \sim 1$
for extremely light masses.

For the exactly massless case, i.e $m = 0$, one sees that the regularized
current~(\ref{Jreg}) diverges due to the $\ln (m / H)$ term, which was
introduced through the procedure of adiabatic subtraction at the
order~$T^{-2}$.
This divergence originates from $\Omega_{\bd{k}}$~(\ref{Omega_k251})
vanishing for $m  = 0$ at
\begin{equation}
 (k_z + e A_z)^2 + k_x^2 + k_y^2 = 0,
\label{k0276}
\end{equation}
and thus blowing up the adiabatic subtraction terms. 
Note that (\ref{k0276}) corresponds to the vanishing of the  physical
momentum  
\begin{equation}
(p_x, \, p_y, \, p_z)
=
\left(
\frac{k_x}{a}, \, 
\frac{k_y}{a}, \, 
\frac{k_z + e A_z}{a} 
\right).
\end{equation}
These observations suggest that the adiabatic expansion taking $ T \to
\infty$ is invalid for zero modes of massless fields, and thus the
method of adiabatic regularization may not be applicable for the
massless case. (See also discussions
in~\cite{Parker:1974qw,Fulling:1974zr}.)
This issue may be resolved by imposing an infrared cutoff on the
momentum integral, 
by considering that in the finite past
the dS expansion started, or the electric field was switched on.

We should also mention that, even if a scalar has a tiny bare mass,
a Hubble-induced mass can be
generated~\cite{Stewart:1994ts,Dine:1995uk}. 
See also discussions on effective mass in dS space in, e.g.,~\cite{Starobinsky:1994bd,Woodard:2006pw,Burgess:2010dd,Serreau:2011fu,Boyanovsky:2012qs}.
The charged scalars need to be protected from mass corrections in order for
a dS universe to actually possess large conductivity under weak electric
fields.

\section{Constraints on Inflationary Magnetogenesis}
\label{sec:magnetogenesis}

In the previous section we studied the production of charged scalars in
a fixed background of a constant electric field and dS expansion.
However the backreaction from the produced scalars may become relevant,
especially when the induced conductivity is huge.
Such considerations are important upon discussing the aftermath of 
the Schwinger process, but can also be used to constrain
electromagnetic fields in a dS universe.
In particular, the backreaction from the Schwinger process can impose
severe constraints on cosmological models for generating primordial
electromagnetic fields during the inflationary dS phase.

The possibility of the cosmological generation of magnetic fields during
the inflationary epoch has been studied by many authors, 
e.g.~\cite{Turner:1987bw,Ratra:1991bn,Bamba:2003av,Demozzi:2009fu,Kanno:2009ei,Fujita:2012rb,Kobayashi:2014sga,Garretson:1992vt,Gasperini:1995dh,Giovannini:2000dj,Davis:2000zp,Anber:2006xt,Martin:2007ue,Bamba:2008ja,Emami:2009vd,Durrer:2010mq,Ferreira:2013sqa,Ferreira:2014hma},
in order to explain the origin of the large-scale magnetic fields in our
universe. 
In such inflationary magnetogenesis scenarios, the magnetic fields are 
generically accompanied by the production of even larger electric fields.
One of the guidelines towards constructing a consistent model has been to
keep the production of electric fields under control in order to avoid
the electric fields from dominating the energy density of the universe~\cite{Bamba:2003av,Demozzi:2009fu,Kanno:2009ei,Fujita:2012rb,Kobayashi:2014sga}.
However, we have seen in the previous section that 
even if the electric fields do not dominate the universe, they can
induce large conductivity in the universe via the Schwinger process,
which may have non-negligible backreaction on the Maxwell fields. 
Therefore, in this section we analyze the effect of the Schwinger process on
magnetogenesis in an inflationary dS spacetime. 
We will see that, unless charged fields are absent in the action,
or are very massive,
the Schwinger effect can spoil the process of magnetogenesis.
Through discussing magnetogenesis, we will also see how the induced
current backreacts on the background Maxwell fields.

\subsection{Model}

The cosmological enhancement of the electromagnetic fields are realized
in inflationary magnetogenesis scenarios by breaking the conformal symmetry
of the Maxwell theory. 
To make our discussions concrete, we focus on a class of models where
the conformal symmetry is broken by a time dependent 
effective coupling~$I$ on the Maxwell kinetic term,\footnote{The
conformal symmetry can be broken alternatively by a mass term for the
photon, $m_{\gamma}^2 A_\mu A^\mu$. In such models,
significant magnetogenesis requires a tachyonic mass, i.e. $m_{\gamma}^2
< 0$, which for example can arise from non-minimal couplings to
gravity~\cite{Turner:1987bw}.
However, such theories have been pointed out to have problems
including the appearance of
a ghost~\cite{Demozzi:2009fu,Dvali:2007ks,Himmetoglu:2009qi}.}
\begin{equation}
 S = \int d^4 x \sqrt{-g} 
\left\{
- \frac{I^2}{4} F_{\mu \nu} F^{\mu \nu}
-g^{\mu \nu} \left( \partial_\mu - i e A_\mu  \right) \varphi^* 
\left( \partial_\nu + i e A_\nu  \right) \varphi
- m^2 \varphi^* \varphi 
\right\}.
\label{actionI^2}
\end{equation}
The coupling~$I$ can be thought of as a function of other degrees of
freedom, such as the inflaton field. 
As in the previous section, we consider Schwinger process with 
a charged complex scalar and
analyze its effect on inflationary magnetogenesis.\footnote{One could
also imagine cases where the effective coupling~$I$ shows up not only in
front of the Maxwell kinetic term, but in front of all the terms in the
Lagrangian. In such cases, a time varying~$I$ can further induce
$\varphi$ production in addition to the Schwinger and gravitational
effects, and therefore the backreaction to the Maxwell fields are
expected to become stronger, resulting in even more stringent
constraints on magnetogenesis than for~(\ref{actionI^2}).}

With the effective coupling, the Maxwell equation is modified to
\begin{equation}
 \nabla^{\nu} \left( I^2 F_{\mu \nu}  \right)  = J_\mu ,
\label{Maxwell-I}
\end{equation}
where the current $J_\mu$ is shown in~(\ref{currentJ}).
In the following, we study the dynamics of the Maxwell
fields in the Coulomb gauge,\footnote{$A_\tau$ cannot be taken to zero
in the presence of charge, however, since we only use the equations
including~$J_\mu$ for obtaining a rough criterion for the
backreaction from~$J_{\mu}$ being non-negligible, we can approximately
set $A_\tau = 0$ in order to simplify the calculations.} 
\begin{equation}
 \partial_i A_i = A_\tau = 0.
\end{equation}
Considering a dS background~(\ref{FRW}), and setting~$I$ to be homogeneous, the
spatial component of the modified Maxwell equation reads
\begin{equation}
 A_i'' - \partial_j \partial_j A_i + 2 \frac{I'}{I}A_i'
= \frac{a^2}{I^2}J_i .
\label{AiEoM}
\end{equation}
We discuss electromagnetic fields defined in terms of~$A_\mu$ (instead of
the normalized $\tilde{A}_\mu = I A_\mu$),
\begin{equation}
E_\mu = u^\nu F_{\mu \nu},
\qquad
 B_\mu = \frac{1}{2} \varepsilon_{\mu \nu \sigma} F^{\nu \sigma},
\end{equation}
because it is $A_\mu$ that the complex scalars couple to with
charge~$e$, and also since 
we consider the coupling~$I$ to be fixed to unity in the present universe. 
Here $u^\mu$ is the $4$-velocity of the comoving observer, and 
$\varepsilon_{\mu \nu \sigma} = \eta_{\mu \nu \sigma \gamma} u^\gamma$,
where $\eta_{\mu \nu \sigma \gamma}$ is a totally antisymmetric
permutation tensor with $\eta_{0123} = -\sqrt{-g} $.
Thus $\varepsilon_{ijk}$ is totally antisymmetric with
$\varepsilon_{xyz} = a^3$.
The time components $B_\tau$ and $E_\tau$ vanish, while the spatial
components are 
\begin{equation}
E_i = -\frac{1}{a} A_i',
\qquad
 B_i = \frac{1}{a^4} \varepsilon_{ijk} \partial_j A_k.
\end{equation}
The magnitude of the fields are 
\begin{gather}
 E^2 \equiv E_\mu E^\mu = \frac{1}{a^2} E_i E_i  = \frac{1}{a^4}A_i'
 A_i' ,
\\
 B^2 \equiv B_\mu B^\mu = \frac{1}{a^2}B_i B_i = \frac{1}{a^4}
\left(\partial_i A_j \partial_i A_j - \partial_i A_j \partial_j A_i
\right).
\label{B^2mag}
\end{gather} 
Then, with the conductivity
\begin{equation}
 \sigma = \frac{J_i}{E_i},
\label{sigma3.9}
\end{equation}
(here we do not take the sum over~$i$ in the right hand side of (\ref{sigma3.9})),
the Maxwell equation~(\ref{AiEoM}) is rewritten as
\begin{equation}
 A_i'' 
- \partial_j \partial_j A_i 
+ \left(  \frac{2 I'}{I} + \frac{a \sigma}{I^2}    \right) A_i'
= 0.
\label{EoMofAi}
\end{equation}

\subsection{In the Absence of Charged Fields}
\label{subsec:absence}

We shall first discuss the idealized situation where any charged fields
are absent in the action.
This subsection will also serve as a brief review of magnetogenesis in
$I^2 FF$ scenarios.

Let us focus on large-scale magnetic fields and neglect the spatial
derivative term in the Maxwell equation~(\ref{EoMofAi}), giving
\begin{equation}
 A_i'' 
+ \frac{2 I'}{I} A_i'
= 0.
\end{equation}
The general solution of this equation is 
\begin{equation}
 A_i = C_1 + C_2 \int  \frac{d \tau }{I^2},
\end{equation}
where $C_1 $ and $C_2$ are constants. For instance, if the
coupling~$I$ decreases in time as
\begin{equation}
 I \propto a^{-s},
\qquad 
\mathrm{with}
\quad
s > \frac{1}{2} ,
\label{Iscaling}
\end{equation}
then the vector potential possesses a growing mode 
\begin{equation}
 A_i \propto a^{2 s-1},
\label{Aipropto}
\end{equation}
and thus the electromagnetic fields are enhanced. 
Hereafter we suppose $I$ to follow the scaling behavior~(\ref{Iscaling})
during inflation, and then stays constant after inflation.\footnote{It could 
also be that the coupling~$I$ approaches a constant 
at some time~$\tau_1$ during inflation,
and thus magnetogenesis terminates before the end of inflation.
In such cases, the constraint on magnetic fields we obtain in~(\ref{B0bound}) 
is modified by $I_{\mathrm{end}} \to I(\tau_1)$,
and also obtains an additional factor of~$a(\tau_1) / a_{\mathrm{end}}$
in the right hand side, which makes the bound more stringent.}  

Focusing on Maxwell fields with a certain wave number~$k$, let us
rewrite the magnetic field amplitude~(\ref{B^2mag}) as\footnote{This
approximate expression of $B^2$ is good enough for obtaining the $E$-$B$
ratio~(\ref{EBratio}). 
More detailed derivations of~(\ref{EBratio}) can be found in the references.}
\begin{equation}
 B^2 \sim \frac{k^2}{a^4} A_i A_i.
\label{B2simk}
\end{equation}
Then, using $A_i' / A_i = (2 s-1) a' / a$ which follows
from~(\ref{Aipropto}), 
the ratio between the electric and magnetic amplitudes with wave
number~$k$ is obtained as
\begin{equation}
 \left|\frac{E}{B} \right| =  (2 s-1)  \frac{ a H_{\mathrm{inf}}}{k}.
\label{EBratio}
\end{equation}
In this section we denote the (nearly) constant Hubble parameter during
inflation by the subscript~``inf''.
As can be seen from~(\ref{B2simk}), the magnetic field after the 
magnetogenesis phase decays as $B \propto a^{-2}$.
Hence we can obtain a relation between the magnetic field strength in the present
universe and the electric field at the end of inflation as
\begin{equation}
\abs{B_0} = \frac{1}{2s-1}
\frac{k}{a_0 H_{\mathrm{inf}}} \frac{a_{\mathrm{end}}}{a_0} \abs{E_{\mathrm{end}}}.
\label{B0Eend317}
\end{equation}
Here the subscript ``$0$'' denotes quantities in the present epoch, and
``end'' at the end of inflation.
We suppose the post-inflationary universe to be first dominated by an
oscillating inflaton, and thus effectively matter-dominated until
reheating happens,
\begin{equation}
 \left( \frac{H_{\mathrm{reh}}}{H_{\mathrm{inf}}} \right)^2 =
 \left( \frac{a_{\mathrm{end}}}{a_{\mathrm{reh}}} \right)^3,
\label{Hrehinf}
\end{equation}
where the subscript ``reh'' denotes quantities at reheating.
After reheating, we consider the entropy to be conserved, i.e. $s
\propto a^{-3}$, and thus obtain\footnote{Upon computing the entropy density
at reheating 
\begin{equation}
s_{\mathrm{reh}} = \frac{2 \pi^2}{45}g_{s*}(T_{\mathrm{reh}})
\left( \frac{90}{\pi^2}\frac{ M_p^2 H_{\mathrm{reh}}^2}{g_*
 (T_{\mathrm{reh}})} \right)^{3/4} ,
\end{equation}
we have chosen the relativistic degrees of freedom to take the 
maximum value allowed in the MSSM, $g_* = g_{s*} = 228.75$.
However we note that this choice affects~(\ref{HrehMp}) only by an
order unity factor; e.g., $g_* = g_{s*} = 10.75$ gives
a factor~$7$ instead of~$6$ in the right hand side.}
\begin{equation}
 \frac{a_{\mathrm{reh}}}{a_0} \approx 6 \times 10^{-32} 
\left( \frac{M_p}{H_{\mathrm{reh}}} \right)^{1/2}.
\label{HrehMp}
\end{equation}
The combination of (\ref{Hrehinf}) and (\ref{HrehMp}) yields the expansion
after inflation,
\begin{equation}
 \frac{a_{\mathrm{end}}}{a_0}
\approx 6 \times  10^{-32}
\left(\frac{H_{\mathrm{reh}}}{H_{\mathrm{inf}}}\right)^{1/6}
 \left(\frac{M_p}{H_{\mathrm{inf}}}\right)^{1/2},
\label{rehinf63}
\end{equation}
which allows us to rewrite (\ref{B0Eend317}) as
\begin{equation}
\frac{\abs{B_0}}{M_p^2}
\approx \frac{6 \times 10^{-32}}{2s-1}
\frac{k}{a_0 M_p }
\left( \frac{H_{\mathrm{reh}}}{H_{\mathrm{inf}}} \right)^{1/6}
\left( \frac{H_{\mathrm{inf}}}{M_p} \right)^{1/2}
\frac{\abs{E_{\mathrm{end}}}}{H_{\mathrm{inf}}^2}.
\label{B-E322}
\end{equation}

\subsection{Constraints from Schwinger Effect}

Let us now study how the above picture of magnetogenesis 
is modified under the existence of
charged scalars in the action. 
From the Maxwell equation~(\ref{EoMofAi}), one can read off the criterion
for the induced current to be negligible as
\begin{equation}
\left| \sigma 
\left( \frac{I'}{a I} \right)^{-1}
\right|
\ll I^2 .
\label{critI}
\end{equation}
In other words, when the ratio between $\sigma$ and the 
rate of magnetogenesis is larger than $\sim I^2$,
the process of magnetogenesis can be strongly affected by the produced scalars. 
The $\sigma$~term in~(\ref{EoMofAi}) is seen to be a friction term for
$A_i'$, thus a positive and large~$\sigma$ decays away the electric
fields and prevents any further magnetogenesis.
For the scaling (\ref{Iscaling}) under consideration, the
criterion~(\ref{critI}) becomes 
\begin{equation}
 \frac{\abs{\sigma}}{H_{\mathrm{inf}}} \ll s I^2.
\label{crits}
\end{equation}
The threshold~$I^2$ in the right hand sides can be understood from the fact that, 
when absorbing the coupling by $\tilde{A}_\mu = I A_\mu$, the
effective charge of~$\varphi$ becomes~$e/I$. 
Hence a smaller~$I$ enhances the backreaction of the produced~$\varphi$
on the Maxwell fields.
This provides an explicit example 
of a problem that arises when $I$ is tiny, often referred to in the
literature as the strong coupling
problem~\cite{Demozzi:2009fu,Gasperini:1995dh}.  
However, even if~$I$ is never smaller than unity, 
we will see that the criterion~(\ref{crits}) severely constrains
inflationary magnetogenesis.

\vspace{\baselineskip}

The criterion~(\ref{crits}) can be used to set an upper bound on the magnetic fields
that can be generated during inflation.
We evaluate the bound by using the approximate expression~(\ref{expMsupp}) for the 
current under strong electric fields (the validity of using this
approximation will shortly be discussed).
With the approximation, the conductivity is expressed as
\begin{equation}
 \frac{\sigma }{H_{\mathrm{inf}}}
\simeq 
\frac{1}{12 \pi^3}
\frac{\abs{e^3 E}}{ H_{\mathrm{inf}}^2}
\exp \left( - \frac{\pi  m^2}{\abs{eE}}  \right) ,
\end{equation}
which is solved for the electric field as
\begin{equation}
 \frac{\abs{E}}{H_{\mathrm{inf}}^2} \simeq
\frac{12 \pi^3}{\abs{e}^3}
\frac{\sigma }{H_{\mathrm{inf}}}
\exp
\left\{
W\left(  
\frac{e^2}{12 \pi^2}
\frac{m^2}{H_{\mathrm{inf}}^2}
\frac{H_{\mathrm{inf}}}{\sigma }
\right)
\right\}.
\label{EforW}
\end{equation}
Here, $W(x)$ is the Lambert $W$-function which is the solution of $W
e^{W} = x$. 
For $x \geq 0$, $W(x)$ is a non-negative and increasing function.
Thus we note that $\abs{E} /
H_{\mathrm{inf}}^2$ is an increasing function of~$\sigma /
H_{\mathrm{inf}}$ when the other parameters are fixed. 
Moreover, since $W(x) \simeq x $ for $0 \leq x \ll 1$, 
the exponential factor in~(\ref{EforW}) approaches unity for small~$m$.
This limit can also be obtained directly by solving the
approximation~(\ref{largeG_sigma}). 

\begin{figure}[htbp]
  \begin{center}
  \begin{center}
  \includegraphics[width=0.5\linewidth]{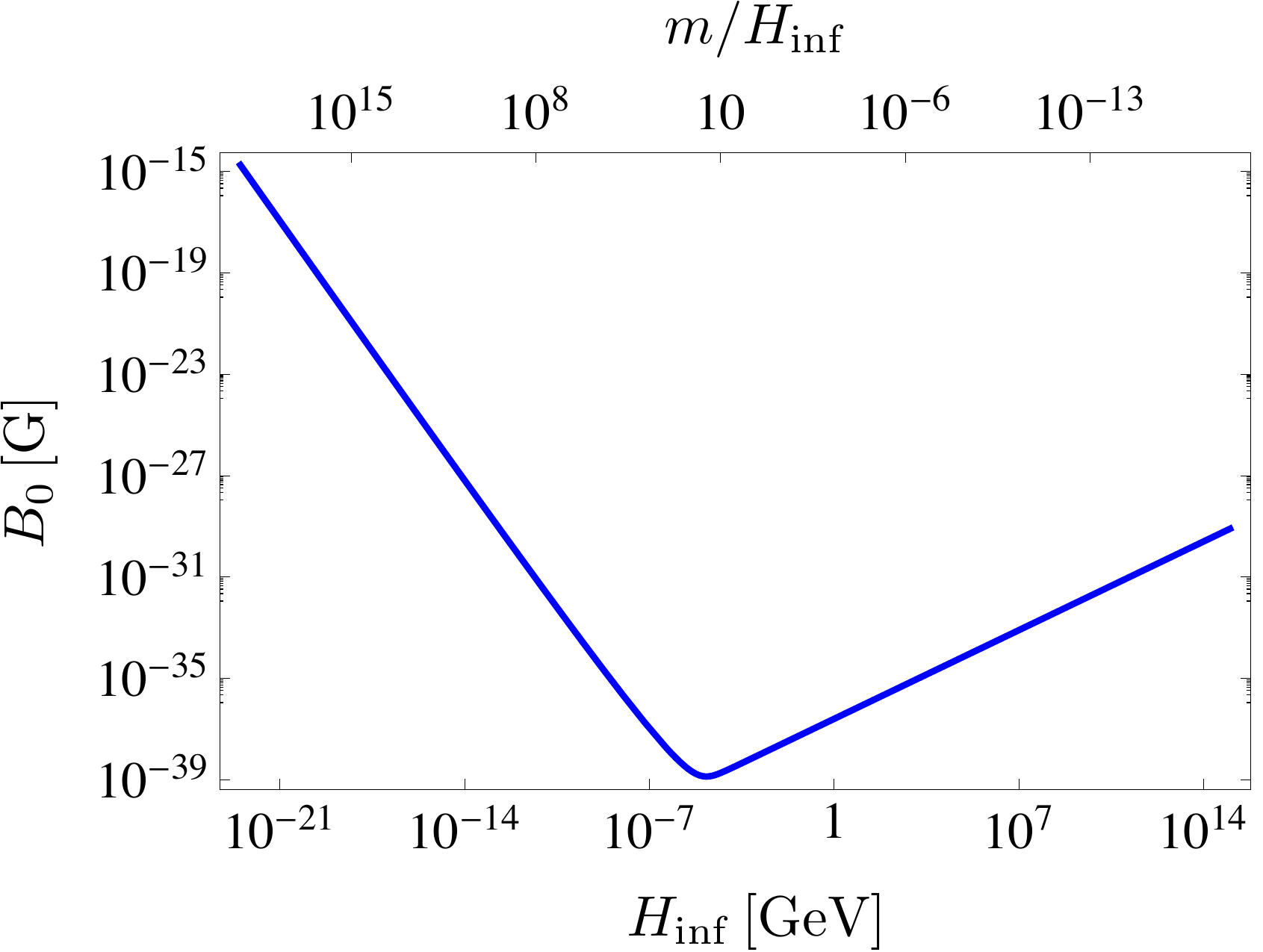}
  \end{center}
  \caption{Upper bound~(\ref{B0bound}) on the present amplitude of
   magnetic fields with
   correlation length $k/a_0 = 1\, \mathrm{Mpc}^{-1}$ from the Schwinger
   effect constraint. A scalar with charge $e^2 = 4 \pi \alpha$ and mass
   $m = 0.5\, \mathrm{MeV}$ is assumed. The effective coupling is set to
   $I_{\mathrm{end}} = 1$. The horizontal axes show the inflation
   scale~$H_{\mathrm{inf}}$ (lower) and the ratio $m / H_{\mathrm{inf}}$
   (upper). In the region where $ m /H_{\mathrm{inf}}$ is
   tiny, further constraints may arise due to the strong enhancement of
   the conductivity at the early stage of magnetogenesis (see the main
   text for details).}
  \label{fig:B-bound0}
  \end{center}
\end{figure}

From (\ref{EforW}), the criterion~(\ref{crits}) is translated into an 
upper bound on the electric field during inflation, 
\begin{equation}
\frac{\abs{E}}{H_{\mathrm{inf}}^2} \lesssim
  \frac{12 \pi^3 s I^2   }{\abs{e}^3}
\exp \left\{W
\left(
\frac{e^2}{12 \pi^2 s I^2}
\frac{m^2}{H_{\mathrm{inf}}^2}
\right)
\right\}, 
\label{EboundINF}
\end{equation}
which imposes a bound on the present magnetic field amplitude via~(\ref{B-E322}),
\begin{equation}
\frac{\abs{B_0}}{M_p^2}
\lesssim
10^{-29} 
\frac{2s}{2s-1}
\frac{k}{a_0 M_p}
\left( \frac{H_{\mathrm{reh}}}{H_{\mathrm{inf}}} \right)^{1/6}
\left( \frac{H_{\mathrm{inf}}}{M_p} \right)^{1/2}
\frac{ I_{\mathrm{end}}^2}{\abs{e}^3}
 \exp \left\{
W
\left(
\frac{e^2}{ 12 \pi^2 s I_{\mathrm{end}}^2}\frac{m^2}{H_{\mathrm{inf}}^2}
\right)
\right\}.
\label{Bbound325}
\end{equation}
Here, $I_{\mathrm{end}}$ denotes the value of the effective coupling at
the end of inflation.
Note from~(\ref{Aipropto}) that $s$ should be larger than 
(and not so close to) $1/2$
for the Maxwell fields to be significantly enhanced during inflation.
Thus $\frac{2s}{2s-1}$ should be of order unity for an efficient magnetogenesis.
Further noting $H_{\mathrm{reh}} \leq H_{\mathrm{inf}}$, 
then the bound can be rewritten as
\begin{equation}
 \abs{B_0} \lesssim
 10^{-28} \mathrm{G} \, 
\left( \frac{k}{a_0}\, \mathrm{Mpc} \right)
\left( \frac{H_{\mathrm{inf}}}{M_p} \right)^{1/2}
\left(\frac{ \sqrt{4 \pi \alpha }  }{\abs{e}}\right)^3
I_{\mathrm{end}}^2
\mathcal{Q},
\label{B0bound}
\end{equation}
where $\mathcal{Q}$ represents the mass dependence,
\begin{equation}
\mathcal{Q} 
= \exp \left\{ W
\left(
\frac{e^2}{ 12 \pi^2 s I_{\mathrm{end}}^2}\frac{m^2}{H_{\mathrm{inf}}^2}
\right)
\right\}
\simeq
  \exp \left\{W
\left(
 10^{-3} \, 
\frac{e^2}{4 \pi \alpha }
\frac{1}{s I_{\mathrm{end}}^2}
\frac{m^2}{H_{\mathrm{inf}}^2}
\right)
\right\}.
\label{Qfactor}
\end{equation}
Here we are using the Heaviside-Lorentz units, thus 
$1\,  G \approx 2 \times 10^{-20} \, \mathrm{GeV}^2 $, and the
elementary charge is $\sqrt{4 \pi \alpha} \approx 0.3$.
We also note $1\, \mathrm{Mpc} \approx 2 \times 10^{29}\,
\mathrm{eV}^{-1}  $.
The mass dependent factor~$\mathcal{Q}$ is a growing
function of $m / H_{\mathrm{inf}}$, with $\lim_{m \to 0} \mathcal{Q} = 1$. 
Moreover, $\mathcal{Q}$ is of order unity when the
argument of~$W$ is smaller than~$\sim 1$. In other words,
the bound~(\ref{B0bound}) is independent of the scalar mass if the mass
ratio satisfies
\begin{equation}
\frac{m^2}{H_{\mathrm{inf}}^2}
 \lesssim 10^3 \, 
\frac{4 \pi \alpha }{e^2} s I_{\mathrm{end}}^2.
\label{Qirrelevant}
\end{equation}
When fixing all the parameters except for~$H_{\mathrm{inf}}$, then the
magnetic upper bound~(\ref{B0bound}) is an increasing function
of~$H_{\mathrm{inf}}$ while (\ref{Qirrelevant}) is satisfied,
scaling as~$H_{\mathrm{inf}}^{1/2}$.
On the other hand, when (\ref{Qirrelevant}) is violated,
the $\mathcal{Q}$ factor becomes important and 
the upper bound turns into a decreasing function of~$H_{\mathrm{inf}}$.
In Figure~\ref{fig:B-bound0} we plot the upper bound~(\ref{B0bound}) as
a function of~$H_{\mathrm{inf}}$, for a fixed set of parameters $k/a_0 =
1\, \mathrm{Mpc}$, $e^2 = 4 \pi \alpha$, $m = 0.5\, \mathrm{MeV}$,
and $I_{\mathrm{end}} = 1$. The scaling factor~$s$ is taken to be of
order unity (its explicit value is unimportant here as varying $s$ by
order unity makes little difference in the plot).
It is clearly seen that the upper bound switches from a decreasing to an
increasing function of~$H_{\mathrm{inf}}$, as the ratio $m /
H_{\mathrm{inf}}$ decreases and starts to satisfy (\ref{Qirrelevant}).

As we have seen, the magnetic field bound~(\ref{B0bound}) derive from 
the Schwinger effect constraint on the electric field~(\ref{EboundINF}),
which can be recast in the form of 
\begin{equation}
 \frac{\abs{e E_{\mathrm{end}}}}{H_{\mathrm{inf}}^2}
\lesssim 10^3\, 
\frac{4 \pi \alpha }{e^2} s I_{\mathrm{end}}^2
\mathcal{Q}.
\label{eE4000}
\end{equation}
Hence for $e^2 \sim 4 \pi \alpha$ and $I_{\mathrm{end}} \sim 1$, 
it is evident that the restrictions on magnetogenesis arise from the Schwinger effect 
in the large~$\abs{\lambda}$ regime, as was assumed upon using the
approximation~(\ref{expMsupp}).
Here we remark that a large charge $e^2 \gg 4 \pi \alpha$ can push the
constrained region into the small~$\abs{\lambda}$ regime and thus
invalidate the usage of~(\ref{expMsupp}). However, the
conductivity in the $\abs{\lambda} \ll 1$ regime is generically larger than
an extrapolation of~(\ref{expMsupp}), and thus the bound is expected
to be even stronger than the form of~(\ref{B0bound}). 
We also note that the approximation~(\ref{expMsupp}) is not necessarily
valid throughout the $\abs{\lambda} \gg 1$ regime; 
it is seen through the purple line in Figure~\ref{fig-sigma} that,
as one moves towards smaller $\abs{\lambda}$, the conductivity $\sigma$
eventually deviates from the exponential fall so that it can smoothly
connect to the plateau at $\abs{\lambda} \ll 1$.
Such behavior also enhances~$\sigma$ relative to that predicted
by~(\ref{expMsupp}), and thus gives a magnetic bound stronger
than~(\ref{B0bound}). 
The situation becomes more severe for cases with extremely light masses
(i.e. $m \ll H_{\mathrm{inf}}$), where 
the conductivity is strongly enhanced at $\abs{\lambda} \ll 1$. 
The large conductivity under weak electric fields can affect
magnetogenesis at its early stage, 
long before the electromagnetic fields grow to values constrained
by the bound~(\ref{B0bound}) 
(though the constraint would also depend on the value of~$I$ during
inflation).\footnote{We could also say that when $m \ll
H_{\mathrm{inf}}$, independently of the magnetogenesis mechanism,
the inflationary universe cannot leave behind arbitrarily small electromagnetic
fields; the strong enhancement of the conductivity under weak electric
fields forbids the electromagnetic fields from existing as a stable
background. In this sense, a ``lower bound'' on the electromagnetic fields
exists for inflationary magnetogenesis with light charged particles.} 
Therefore the plot in Figure~\ref{fig:B-bound0} should be considered as
a conservative bound for regions where the ratio $m / H_{\mathrm{inf}}$
is tiny.

We should also comment on the applicability of the results from the 
previous section, where we considered a constant and uniform electric
field, on inflationary magnetogenesis where the electromagnetic fields
with finite correlation lengths are continuously being produced.
Here it should be noted that the typical time scale for the enhancement
of the electromagnetic fields is, in the case~(\ref{Aipropto}) under
consideration, of order the Hubble time~$H^{-1}$.
Moreover, the electromagnetic fields are significantly enhanced after
exiting the Hubble horizon, and thus we have given
constraints on wave modes that are sufficiently larger than the
horizon at the end of the magnetogenesis phase.
On the other hand, the constraints are mostly due to the Schwinger
process in the $\abs{e E} \gg H_{\mathrm{inf}}^2 $ regime,
and thus the produced charged scalars typically have wave modes much
smaller than the Hubble radius, cf.~(\ref{tau-k-est}).
In this regime, it could also be checked that the $\varphi$~population
is always dominated by the particles newly created within a Hubble time. 
Thus the length and time scales relevant to the Schwinger process do
not exceed those of the electric fields, 
validating our procedure of modeling the electric fields produced during
inflation as being
constant and uniform upon evaluating the Schwinger effect
constraints.

\vspace{\baselineskip}

\begin{figure}[htbp]
  \begin{center}
  \begin{center}
  \includegraphics[width=0.55\linewidth]{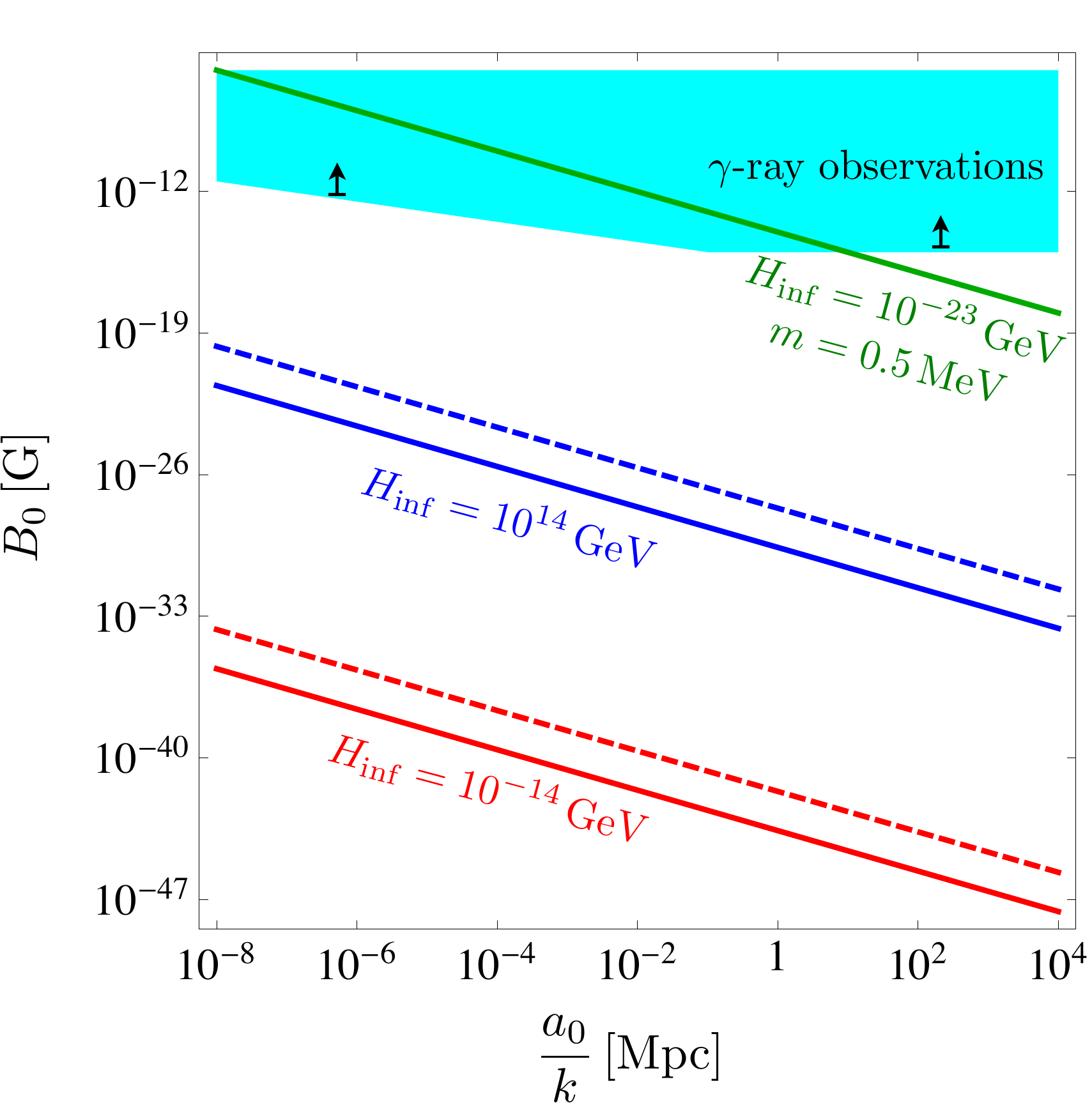}
  \end{center}
  \caption{Upper bound on the present amplitude of magnetic fields
   produced during inflation from 
   Schwinger effect constraints. The blue and red lines are,
   respectively, for $H_{\mathrm{inf}} =  10^{14} \, \mathrm{GeV}$ and
   $H_{\mathrm{inf}} = 10^{-14} \, \mathrm{GeV}$, 
   with $ m \lesssim  10 H_{\mathrm{inf}}$ (solid lines)
   and $ m =  10^3  H_{\mathrm{inf}}$ (dashed lines).
   The green line shows the case for the lowest possible inflation scale
   $H_{\mathrm{inf}} = 10^{-23} \, 
   \mathrm{GeV}$, with $m = 0.5\, \mathrm{MeV}$. The charge of
   the complex scalar is fixed to $e^2 = 4 \pi \alpha$, and the
   effective coupling to $I_{\mathrm{end}} = 1$.
The cyan shaded region shows the magnitude of intergalactic magnetic
   fields suggested by gamma-ray observations, whose lower bound is from~\cite{Taylor:2011bn}.}
  \label{fig:B-bound}
  \end{center}
\end{figure}

In Figure~\ref{fig:B-bound} we plot the upper bound on the amplitude of
the magnetic fields in the present universe~(\ref{B0bound}) as a
function of the correlation length. The charge of the complex scalar 
is set to the elementary charge $e^2 = 4 \pi \alpha$, and the coupling
at the end of inflation to~$I_{\mathrm{end}} = 1$.
The scaling factor~$s$ is chosen to be of order unity; the bound depends
sufficiently weakly on~$s$ such that its explicit value is not important
here. 
For the chosen sets of parameters in the figure,
magnetogenesis is constrained by the Schwinger effect in the
large~$\abs{\lambda}$ regime where the approximation~(\ref{expMsupp})
is valid, and thus the magnetic bounds can be fully
described by the expression~(\ref{B0bound}).

Each line represents the upper bound for a different set of the inflation
scale~$H_{\mathrm{inf}}$ and the scalar mass~$m$.
The case of a high-scale inflation with $H_{\mathrm{inf}} = 10^{14} \,
\mathrm{GeV}$ (corresponding to a tensor-to-scalar ratio of $r \simeq
0.2$, as recently suggested in~\cite{Ade:2014xna}) is shown as blue lines;
the solid line is for $m \lesssim 10 H_{\mathrm{inf}}$,
and the dashed line for $m = 10^3 H_{\mathrm{inf}}$.
In the former case, the mass dependent
factor~(\ref{Qfactor}) is $\mathcal{Q}\sim 1$, while for the latter
$\mathcal{Q} \sim 10^2$ and thus the bound is relaxed.

The bounds should be compared with results from recent gamma ray
observations, that suggest the existence of intergalactic magnetic fields
of strength
\begin{equation}
 \abs{B_0} \gtrsim 10^{-15}\, \mathrm{G},
\label{blazar}
\end{equation}
when the correlation length is of $\mathrm{Mpc}$ scales or
larger~\cite{Tavecchio:2010mk,Neronov:1900zz,Ando:2010rb,Taylor:2011bn,Takahashi:2013uoa,Finke:2013bua,Tashiro:2013ita}.
If the correlation length~$\lambda_B$ is much smaller than a Mpc, the lower bound
improves as $\lambda_B^{-1/2}$.
The suggested magnetic field strength is shown in the figure as the
cyan shaded region, where the lower bound is taken
from~\cite{Taylor:2011bn} (for the case of extended cascade emission).
The observational bound has astrophysical uncertainties 
(see e.g.~\cite{Broderick:2011av,Miniati:2012ge})
and more detailed work will be required to verify the claim,
however we already see from 
the plot that the Schwinger effect severely constrains 
the inflationary magnetogenesis scenario from producing such large-scale
magnetic fields; the blue solid line shows an upper bound of $\abs{B_0}
\lesssim 10^{-30}\, \mathrm{G}$ on Mpc scale and beyond.
For other observational constraints on magnetic fields, see the
review~\cite{Durrer:2013pga}.

Here we should note that, in order to generate magnetic power of (\ref{blazar}) on
Mpc scales from the magnetogenesis with the scaling behavior~(\ref{Iscaling}),
then the inflation scale should actually satisfy 
$H_{\mathrm{inf}} < 10^{-32} M_p \sim 10^{-14} \, \mathrm{GeV}$,
otherwise the produced electric fields end up dominating the energy
density of the universe, as derived in~\cite{Kobayashi:2014sga}. (See
also~\cite{Bamba:2003av,Demozzi:2009fu,Fujita:2012rb}.)\footnote{Combining 
(\ref{B-E322}) with the 
requirement that the electric field should not dominate 
the energy density of the universe, i.e.,
\begin{equation}
 \rho_{\mathrm{E}} \sim \frac{I^2}{2} E^2 < 3 M_p^2 H_{\mathrm{inf}}^2,
\end{equation}
and further demanding
$I_{\mathrm{end}}^2 \gtrsim 1$ to avoid strong couplings in the theory,
one can check that $H_{\mathrm{inf}} \lesssim 10^{-32} M_p$ is necessary
for magnetic fields of $\abs{B_0} \gtrsim 10^{-15}\, \mathrm{G}$
to be produced on scales of Mpc or larger.}
Thus in the figure we also plot the magnetic bound for $H_{\mathrm{inf}}
= 10^{-14} \, \mathrm{GeV}$ as red lines, again with the solid line for $m \lesssim 10
H_{\mathrm{inf}}$ and the dashed line for $m = 10^3 H_{\mathrm{inf}}$.
With such a low inflation scale, the red solid line now gives $\abs{B_0} \lesssim
10^{-44}\, \mathrm{G}$ on Mpc scales.

As long as $ m \lesssim 10 H_{\mathrm{inf}}$,
the bound scales as~$H_{\mathrm{inf}}^{1/2}$ and thus becomes
more severe for lower inflation scales.
However, the situation is
different when the mass is much larger than~$H_{\mathrm{inf}}$.
We have seen in Figure~\ref{fig:B-bound0} that,
when the mass is sufficiently large such that (\ref{Qirrelevant}) is
violated, then the upper bound~(\ref{B0bound}) turns into a decreasing function
of~$H_{\mathrm{inf}}$.
Thus we further plot the case of a charged field with an electron mass
$m = 0.5\, \mathrm{MeV}$, and an extremely low-scale inflation
$H_{\mathrm{inf}} = 10^{-23} \, \mathrm{GeV}$, which is the lowest
possible scale compatible with Big Bang
Nucleosynthesis~\cite{Kawasaki:2000en,Hannestad:2004px},
although it would require instantaneous reheating and baryogenesis.
This extreme case with $m / H_{\mathrm{inf}} = 5 \times 10^{19}$ is
shown as the green line in the plot. Due to the very large mass-Hubble
ratio, the conductivity is suppressed and thus relaxes the bound
by $\mathcal{Q} \sim 10^{34}$. In particular, the upper bound on the Mpc scale
is $\abs{B_0} \lesssim 10^{-14}\, \mathrm{G}$, which is
comparable to the value of the observational lower bound~(\ref{blazar}). 

\vspace{\baselineskip}

In summary, under the existence of charged fields in the action, 
the Schwinger effect introduces a serious obstacle towards inflationary
magnetogenesis. This is manifested in the form of an upper
bound~(\ref{B0bound}) on the produced magnetic field, with
possible corrections which typically make the bound more severe,
as discussed below~(\ref{eE4000}). 
If for example the charged field has a mass of order the Hubble scale or
smaller, and carries charge of order the elementary charge, then the
Schwinger constraint eliminates the possibility of 
the discussed inflationary magnetogenesis scenario being
responsible for producing the 
extragalactic magnetic fields~(\ref{blazar}) suggested by gamma ray
observations.
For cases with extremely light charged fields, i.e. $m \ll H_{\mathrm{inf}}$,
the phase of magnetogenesis may not be able to even start,
as the induced conductivity under weak electric fields is substantially enhanced.
We stress that, as long as the mass of the charged field is not much
larger than the Hubble scale, the constraint is more severe for
lower inflation scales. Hence lowering the energy scale of inflation 
(as was considered in~\cite{Ferreira:2013sqa,Ferreira:2014hma} to circumvent previous
constraints) does not improve the situation. 

The Schwinger constraint can be relaxed if all charged
fields have tiny charges, or if their masses are much larger than the inflationary
Hubble scale. 
For instance, the running of the charge at high energies may suppress the
charges during inflation. 
Alternatively, if inflation is driven by the Higgs
field~\cite{Bezrukov:2007ep}, then charged particles 
in the Standard Model may acquire very large masses
while the Higgs field takes large field values.
Another possibility of relaxing the constraint is 
to have a large value for the effective coupling~$I$ at the end of the
magnetogenesis phase, so that the backreaction from the induced current
is suppressed. 
(We repeat that $I$ here denotes the 
relative factor between the Maxwell kinetic term and the 
coupling term between the vector potential and the charged field.
Thus when absorbing $I$ into the 
definition of~$A_{\mu}$, then a large~$I$ corresponds to a tiny
effective charge.)
If $I$ is to approach unity in the present universe, then 
one could imagine a case where $I$ takes a large value at the end of
inflation, and keeps decreasing after inflation.
Such a possibility was investigated in~\cite{Kobayashi:2014sga}, where
magnetogenesis further continues 
in the post-inflationary epoch until reheating.
Constraints from the Schwinger effect can readily be applied for 
post-inflationary mechanisms of magnetogenesis as well;
due to the low Hubble rate, 
the Schwinger process is expected to receive strong mass
suppression, though we defer a careful study of Schwinger effect in
post-inflationary scenarios to a future work.
We should also note that the Schwinger constraint is evaded if the
charged fields are absent from 
the action during inflation, e.g., it could be that the terms for the
charged fields in the action somehow emerge in the post-inflationary universe.

\section{Conclusions}
\label{sec:conc}

In this work, we have analyzed particle creation by electric and
gravitational fields in a four-dimensional dS space. 
In addition to the usual Bogoliubov computations, we 
calculated the expectation value of the induced current. 
By directly evaluating the current, we could investigate regimes where
an adiabatic vacuum does not necessarily exist in the asymptotic future
for the charged particles. 
However, divergences had to be removed from the expectation value of the
current operator. 
To this end, we applied the adiabatic regularization method.
We saw that subtracting terms up to quadratic order in the adiabatic
expansion removes all infinities, while also yields results that
have the correct behavior in the flat space limit.  

The expression for the regularized current is presented in (\ref{Jreg}).
Under strong electric fields $\abs{eE} \gg H^2 $, 
the limiting form of the current is shown 
in~(\ref{largeG_J}) (or (\ref{expMsupp})), whose behavior coincides with
that in flat space. In particular, the linear dependence of the
conductivity on the electric field, i.e. $\sigma \propto E$, 
which is inherent in a four-dimensional space,
plays an important role upon constraining electric fields in the early
universe. 
On the other hand, under weak electric fields~$\abs{eE} \ll H^2 $, the
approximate expression for the induced current is given
in~(\ref{sigma_weakE}).
The conductivity in this regime is independent of the electric field,
and moreover is inversely proportional to the mass squared, 
i.e. $\sigma \sim 10^{-2} \times e^2 H / m^2$.
Thus the dS space acquires a large conductivity under weak electric
fields for small masses, i.e. $m \ll H$.
This intriguing phenomenon with small masses is supported by the
infrared modes of the charged scalar, and was also seen to happen in a 
two-dimensional dS~\cite{Frob:2014zka}.
For massive particles, the contribution to the current from each wave
mode does not grow indefinitely towards the infrared, 
however we note that the scaling $\sigma \propto m^{-2}$ holds for arbitrary mass. 
Thus even massive charged particles can give rise to a non-negligible
conductivity under weak electric fields in four-dimensional dS space. 
This should be contrasted to the case of two-dimensional dS where
$\sigma$ is exponentially suppressed at large masses. 

We remark that the exactly massless case cannot be handled
in the formalism presented in this paper, as the adiabatic expansion
breaks down for the zero modes of the massless field. 
It should also be noted that loop corrections may generate large masses to the
charged scalars, and thus avoid the dS universe from obtaining an
extremely large conductivity under weak electric fields. 
In this paper, we have adopted the method of adiabatic regularization,
however it would be important to compute the current with a different
regularization or renormalization scheme and compare the results. We
leave this for future work. 

In the second half of the paper, we applied the above results to the
early universe in order to constrain cosmological scenarios for
generating large-scale magnetic fields in our universe. 
We showed that the electric fields generated together with the magnetic
fields can induce sufficiently large conductivity to terminate the phase
of magnetogenesis. 
We have especially focused on inflationary magnetogenesis models with a 
modified Maxwell kinetic term $I(t)^2 F_{\mu \nu} F^{\mu \mu}$, whose
coupling scales as $I \propto a^{-s}$.
The main constraints arise from the strong electric field regime
$\abs{eE} \gg H^2$, where the behavior of the Schwinger process is
similar to that in flat space. 
The upper bound from the Schwinger constraint on the produced magnetic
amplitude is given in (\ref{B0bound}), and the bounds at various
length scales are displayed in Figure~\ref{fig:B-bound}. 
For instance, if the charged field has a mass of order the Hubble scale or
smaller, and carries charge of order the elementary charge, then
magnetic fields with correlation length of Mpc or larger is bounded as
$\abs{B_0} \lesssim 10^{-30}\, \mathrm{G}$ for all possible inflation
scales. 
Although the explicit bound depends on the masses and charges of the
particles, we have shown that unless charged fields are absent from
the Lagrangian during inflation, 
the Schwinger effect makes it a formidable task for inflationary
magnetogenesis to produce the extragalactic magnetic fields of
$\sim 10^{-15}\, \mathrm{G}$ suggested by gamma ray observations. 

In this paper we have focused on a certain class of inflationary
magnetogenesis scenarios, however it would be interesting to 
systematically constrain inflationary magnetogenesis in general from
the Schwinger effect. 
It is also important to constrain
non-inflationary mechanisms,
such as the scenario in~\cite{Kobayashi:2014sga} which generates
magnetic fields during the matter-dominated phase prior to reheating, and
\cite{Vachaspati:1991nm,Cornwall:1997ms,Vachaspati:2008pi} during
phase transitions.

Through the discussions on magnetogenesis scenarios, we have seen 
that the Schwinger effect gives rise to strong constraints on
electromagnetic fields under the existence of charged fields in the
action. 
This, in turn, suggests the exciting possibility of extracting 
information about the charged fields in the Lagrangian, from
the (non-)detection of primordial magnetic fields in our universe.

\section*{Acknowledgements}

We would like to thank Jaume Garriga, Shinji Mukohyama, and Tanmay
Vachaspati for useful comments on a draft.
TK is also grateful to Marcelo Alvarez, Tony Chu, Chris Thompson, 
Yuki Watanabe, and Aaron Zimmerman for very helpful discussions. 
This work was supported by the Natural Science and Engineering Research
Council of Canada, the University of Waterloo and Perimeter Institute
for Theoretical Physics. Research at the Perimeter Institute is
supported by the Government of Canada through Industry Canada and by the
Province of Ontario through the Ministry of Research \& Innovation.


\appendix

\section{Some Properties of Whittaker Functions}
\label{app:Whittaker}

In this appendix we lay out some of the properties of the Whittaker
Functions that are useful for the discussions in
Section~\ref{sec:Schwinger}.
For more details, see e.g.~\cite{Olver:2010:NHMF}.

The Whittaker functions
\begin{equation}
\begin{split}
 W_{\kappa, \mu} (z) &= e^{-\frac{z}{2}} z^{\frac{1}{2} + \mu}
U\left( \tfrac{1}{2} + \mu - \kappa, 1 + 2 \mu, z \right),
\\
 M_{\kappa, \mu} (z) &= e^{-\frac{z}{2}} z^{\frac{1}{2} + \mu}
M\left( \tfrac{1}{2} + \mu - \kappa, 1 + 2 \mu, z \right),
\end{split}
\end{equation}
defined in terms of Kummer's confluent hypergeometric functions $U$ and $M$,
are solutions of the differential equation
\begin{equation}
\frac{d^2 \mathcal{W}}{dz^2} + 
\left\{
\frac{1}{z^2}\left( \frac{1}{4} - \mu^2 \right) + \frac{\kappa }{z}
 - \frac{1}{4} 
\right\} \mathcal{W} = 0.
\label{mathcalWeq}
\end{equation}
Here, $M_{\kappa, \mu}(z)$ does not exist when $2 \mu = -1, -2, \cdots$.
The fundamental pairs of solutions of~(\ref{mathcalWeq}) are formed by
$W_{\kappa, \mu}(z)$, $W_{-\kappa, \mu}(e^{\pi i} z)$ (for $-\frac{3}{2}
\pi \leq \arg z \leq \frac{1}{2} \pi $), 
or $M_{\kappa, \mu}(z)$, $M_{\kappa, -\mu}(z)$ (for $-\pi \leq \arg z
\leq \pi$ and $2 \mu \neq 0, \pm 1, \pm 2, \, \cdots$).

The functions have the properties
\begin{equation}
(W_{\kappa, \mu}(z))^* = W_{\kappa^*, \mu^*}(z^*),
\quad
(M_{\kappa, \mu}(z))^* = M_{\kappa^*, \mu^*}(z^*),
\end{equation}
\begin{equation}
 W_{\kappa, \mu}(z) = W_{\kappa, - \mu}(z),
\quad
 M_{\kappa, \mu} (z e^{\pm \pi i}) = \pm i e^{\pm \mu \pi i} M_{-\kappa,   \mu} (z) ,
\end{equation}
and are related through the formula (for $2 \mu \neq 0, \pm 1, \pm2, \cdots$):
\begin{equation}
 W_{\kappa, \mu} (z) = \frac{\Gamma(-2 \mu)}{\Gamma(\frac{1}{2} - \mu -
  \kappa) }M_{\kappa, \mu} (z)
 + \frac{\Gamma(2 \mu)}{\Gamma(\frac{1}{2} + \mu - \kappa)} M_{\kappa,
 -\mu} (z).
\label{formulaWM}
\end{equation}
The Wronskians are
\begin{equation}
\begin{split}
 W_{\kappa, \mu} (z) \frac{d W_{-\kappa, \mu } (e^{\pm \pi i } z)}{dz}
&- \frac{d W_{\kappa, \mu} (z) }{dz} W_{-\kappa, \mu } (e^{\pm \pi i} z)
= e^{\mp \kappa \pi i },
\\
 M_{\kappa, \mu} (z) \frac{d M_{\kappa, -\mu } (z)}{dz}
&- \frac{d M_{\kappa, \mu} (z) }{dz} M_{\kappa, -\mu } (z)
= -2 \mu.
\end{split}
\end{equation}
As $\abs{z} \to \infty$, the function $W_{\kappa, \mu} (z)$ has a limiting form of 
\begin{equation}
 W_{\kappa, \mu} (z) \sim e^{-z/2} z^\kappa,
\qquad
\mathrm{for}
\, \, \, 
\abs{\arg z} < \tfrac{3}{2} \pi .
\label{Wlimitzinf}
\end{equation}
As $z \to 0$, the function $M_{\kappa, \mu} (z)$ approaches
\begin{equation}
 M_{\kappa, \mu} (z ) \sim z^{\mu + 1/2}.
\label{Mlimitz0}
\end{equation}

We note that, in this paper we take the principal values $-\pi \leq
\arg \varpi \leq \pi $  for the phase of complex numbers~$\varpi$.

\section{Computation of the Current Before Regularization}
\label{app:int}

In this appendix we perform the three-dimensional momentum integral
in~(\ref{jzWkmu}) 
for obtaining the expectation value of the current before its
divergences are regularized.
We follow the calculational procedure in~\cite{Frob:2014zka} for a
one-dimensional momentum integral, with some modifications along the way.

The integral under consideration is
\begin{multline}
 \int d^3 k\, \left(k_z + e A_z \right)
\frac{e^{i \kappa \pi}}{2 k}
\left| W_{\kappa, \mu} (z) \right|^2 
=
2 \pi 
\int^\infty_0 dk\, k^2
\int^1_{-1} d r\, 
\left( kr + \frac{\lambda }{\tau }  \right)
\frac{e^{r \lambda \pi }}{2 k}
\left| W_{-i r \lambda , \mu}  (2 k i \tau ) \right|^2
\\
 =
 -\frac{\pi }{\tau^3}
\lim_{\xi \to \infty}
\int^{\xi }_0 d v\, v
\int^1_{-1} dr\, \left( r v - \lambda  \right)
e^{r \lambda  \pi} \left| W_{-i r \lambda, \mu} (-2 i v) \right|^2,
\label{expA1}
\end{multline}
where we introduced real variables:
\begin{equation}
 r = \frac{k_z}{k},
\qquad
\lambda = \frac{e E}{H^2},
\qquad
 v = - k \tau .
\end{equation}
We have also put a cutoff~$\xi$ on the $v$-integral, which we will take
to infinity at the end of the calculation.
Note from the definition (\ref{zkappamu}) that $\mu$ is either real or
purely imaginary, and that its real part lies in the range of
$0 \leq \abs{\mathrm{Re} (\mu)} \leq \frac{3}{2} $.
In the following analyses we further suppose
\begin{equation}
 \mu^2 \neq 0, \, \frac{1}{4},\, 1, \, \frac{9}{4} ,
\label{muconst}
\end{equation}
for calculational convenience.
The excluded cases can be approached by taking the limits
$\mu^2 \to 0, \frac{1}{4}, 1, \frac{9}{4} $ of the final result.

Let us rewrite the Whittaker function using the 
Mellin--Barnes integral representation that is valid
for $\frac{1}{2} \pm \mu - \kappa \neq 0, -1, -2, \ldots,$
and $\abs{\arg z} < \frac{3}{2} \pi $
(recall that in this paper we take the principal values $-\pi
\leq \arg \varpi \leq \pi $ for the phase of complex numbers):
\begin{equation}
 W_{\kappa, \mu} (z) = \frac{e^{-z/2}}{2 \pi i}\int^{i \infty}_{-i
  \infty}
\frac{\Gamma (\frac{1}{2} + \mu + s) \Gamma (\frac{1}{2} - \mu + s)
\Gamma (-\kappa - s)}{\Gamma (\frac{1}{2} + \mu - \kappa ) \Gamma
(\frac{1}{2} - \mu - \kappa )}
z^{-s} ds,
\label{MellinBarnes}
\end{equation}
where the contour of integration separates the poles of
$\Gamma (\frac{1}{2} + \mu + s) \Gamma (\frac{1}{2} - \mu + s)$ from
those of $\Gamma (-\kappa - s)$.
Also using $( W_{-i r \lambda, \mu} (-2 i v) )^* = W_{i r \lambda,
\mu^*} (2 i v)$, then (\ref{expA1}) can be rewritten as
\begin{equation}
\begin{split}
 -\frac{\pi }{\tau^3}
&\lim_{\xi \to \infty}
\int^{\xi}_0 dv\, v 
\int^1_{-1} dr\, (rv - \lambda) e^{r \lambda \pi }
\\
&\times
 \int^{i \infty}_{-i \infty} \frac{ds}{2 \pi i}
\frac{\Gamma (\frac{1}{2} + \mu + s) \Gamma (\frac{1}{2} - \mu + s)
\Gamma (i r \lambda - s)}{ \Gamma (\frac{1}{2} + \mu + i r \lambda)
\Gamma (\frac{1}{2} - \mu + i r \lambda)} (-2 i v)^{-s} 
\\
& \times
 \int^{i \infty}_{-i \infty} \frac{dt}{2 \pi i}
\frac{\Gamma (\frac{1}{2} + \mu^* + t) \Gamma (\frac{1}{2} - \mu^* + t)
\Gamma (-i r \lambda - t)}{ \Gamma (\frac{1}{2} + \mu^* - i r \lambda)
\Gamma (\frac{1}{2} - \mu^* - i r \lambda)} (2 i v)^{-t} .
\label{expA4}
\end{split}
\end{equation}
Note that the $\mu^*$'s in the expression can be converted into $\mu$'s, since
$\mu^*$ is equal to either $\mu$ or $-\mu$. 

The integration contours of $s$ and $t$ are arbitrary, as long as they
separate the poles as discussed below~(\ref{MellinBarnes}),
and run from minus to plus infinity in the imaginary direction.
Therefore, we choose the contours to always satisfy
\begin{equation}
 \mathrm{Re}(s), \, \mathrm{Re}(t) < 1.
\label{maru-ro}
\end{equation}
Then the $v$-integral in (\ref{expA4}) can be carried out as
\begin{equation}
\begin{split}
 -\frac{\pi }{\tau^3}
\lim_{\xi \to \infty}
& \int^{1}_{-1} dr
\frac{e^{r \lambda \pi}}{\Gamma (\frac{1}{2} + \mu + i r \lambda)
\Gamma (\frac{1}{2} - \mu + i r \lambda)
\Gamma (\frac{1}{2} + \mu - i r \lambda)
\Gamma (\frac{1}{2} - \mu - i r \lambda)
}
\\
& \times
 \int^{i\infty}_{-i \infty} \frac{ds}{2 \pi i} \, 
\Gamma \left(\tfrac{1}{2} + \mu + s\right) \Gamma \left(\tfrac{1}{2} - \mu
	+ s\right) \Gamma (i r \lambda - s)
\int^{i \infty}_{-i \infty}\frac{dt}{2 \pi i} \,  f_{r,s}(t),
\label{expA6}
\end{split}
\end{equation}
where 
\begin{equation}
\begin{split}
f_{r,s}(t)  
&=
\Gamma \left(\tfrac{1}{2} + \mu + t\right) \Gamma \left(\tfrac{1}{2} - \mu
	+ t\right) \Gamma (- i r \lambda - t)
\\
& \qquad \qquad 
\times
\frac{1}{4}
 e^{i \frac{\pi }{2} (s-t)}
(2 \xi)^{2-s-t}
\left( \frac{r \xi}{3-s-t} - \frac{\lambda }{2-s-t}\right).
\end{split}
\end{equation}
For a fixed set of $r$ and $s$, the function $f_{r,s}(t) $ can have
singularities at $t = -\frac{1}{2} \pm \mu - n$ (where $n = 0, 1, 2,
\cdots$),
located on the left side of the integration contour of~$t$, 
and $t = - i r \lambda + n, \, 2-s, \, 3-s$, on the right side of the contour.

Upon integrating~$f_{r,s}(t)$ over~$t$, let us further specify the integration
contour of~$s$ by requiring  
\begin{equation}
 -1 < \mathrm{Re} (s)
\label{maru-ha}
\end{equation}
to be always satisfied.
We then carry out the $t$-integral by closing its contour in the right
half-plane, without passing through any of the poles. 
The added integration contour of~$t$ does not contribute to the result,
since an integral of~$f_{r,s}(t)$ over a finite path along the
real direction vanishes 
at $\mathrm{Im} (t) \to \pm \infty$, and also because any integral in
the region $\mathrm{Re}(t) > 4$ vanishes in the limit $\xi \to \infty$ 
due to the condition~(\ref{maru-ha}). 
The residues of~$f_{r,s}(t)$ inside the closed contour also vanish as
$\xi \to \infty$, except for those at the simple poles:
\begin{equation}
  t = -i r \lambda, \, -i r \lambda + 1, \, -i r \lambda +2, \, 
-i r \lambda +3, \, 2-s, \, 3-s.
\end{equation}
Among the six poles, the ones at $t = -i r \lambda, \cdots, -i r \lambda
+ 3$ give residues that have explicit $\xi$-dependence, while the
residues at $t = 2-s, 3-s$ are independent of~$\xi$. 
Instead of showing the full expression, 
in order to reduce clutter we schematically write 
\begin{equation}
 \lim_{\xi \to \infty} \int \frac{dt}{2 \pi i}
  f_{r,s} (t)
=
 \lim_{\xi \to \infty} \mathcal{O}(\xi^{-s+i r \lambda + 3},
\dotsc,
\xi^{-s+i r \lambda })
+ \mathcal{O}(\xi^0).
\label{schematicfrs}
\end{equation}

The $s$-integral in (\ref{expA6}) with the $\xi$-dependent terms
in~(\ref{schematicfrs}) can be
carried out similarly to the $t$-integral; 
closing the contour in the right half-plane, the only residues that survive
as $\xi \to \infty$ are those at 
\begin{equation}
 s = i r \lambda,\, i r \lambda + 1, \, i r \lambda + 2, \, i r \lambda  + 3.
\end{equation}
These poles are not necessarily simple poles, and thus gives an
expression that involves digamma functions $\psi (z) = \Gamma'(z) /
\Gamma (z)$, 
\begin{equation}
\begin{split}
&\lim_{\xi \to \infty}
 \int^{i\infty}_{-i \infty} \frac{ds}{2 \pi i} \, 
\Gamma \left(\tfrac{1}{2} + \mu + s\right) \Gamma \left(\tfrac{1}{2} - \mu
	+ s\right) \Gamma (i r \lambda - s)
\mathcal{O}(
\xi^{-s+i r \lambda + 3},
\dotsc,
\xi^{-s+i r \lambda }
)
\\
& 
= 
e^{-r \lambda \pi} \, 
\Gamma \left(\tfrac{1}{2} + \mu + i r \lambda\right)
\Gamma \left(\tfrac{1}{2} - \mu + i r \lambda\right)
\Gamma \left(\tfrac{1}{2} + \mu - i r \lambda\right)
\Gamma \left(\tfrac{1}{2} - \mu - i r \lambda\right)
\\
& \quad 
\times 
\lim_{\xi \to \infty}
\Biggl[
\frac{r}{3} \xi^3
- \frac{\lambda}{2} (1-r^2)  \xi^2 
-\frac{r}{8} 
\left\{
1 - 4 \mu^2 + 4 (2 - 3 r^2) \lambda^2
\right\} \xi 
\\
& \qquad \qquad \, \, \, \quad
 +   \frac{\lambda }{8}
\left\{
1 - 4 \mu^2 + (-7 - 12 \lambda^2 + 12 \mu^2) r^2 + 20 \lambda^2 r^4
\right\}
\\
& \qquad \qquad \qquad \qquad \quad
 \times \left\{
\ln (2 \xi) - \psi \left(\tfrac{1}{2} + i r \lambda - \mu \right)
- \psi \left( \tfrac{1}{2} + i r \lambda + \mu  \right)
\right\}
+ \dotsb
\Biggr].
\label{temrsfromXi}
\end{split}
\end{equation}
In the final line, we have abbreviated terms that are independent
of~$\xi$ by dots. 

On the other hand, the $s$-integral of the $\xi$-independent terms
in~(\ref{schematicfrs}) can be written as,
\begin{equation}
\begin{split}
&\int^{i\infty}_{-i \infty} \frac{ds}{2 \pi i} \, 
\Gamma \left(\tfrac{1}{2} + \mu + s\right) \Gamma \left(\tfrac{1}{2} - \mu
	+ s\right) \Gamma (i r \lambda - s)
\mathcal{O}( \xi^0 )
\\
& 
=  \int^{i \infty}_{-i \infty} \frac{ds}{2 \pi i}
\frac{e^{i \pi s}}{
\cos\{ \pi(\mu + s) \} \cos \{\pi (-\mu + s)\} \sin \{\pi (s - i r \lambda )\}
}
\\
& \qquad \qquad \qquad \qquad \qquad \qquad \qquad \qquad
\times
\left\{
  \frac{d_r  }{s-i r \lambda }
+ g_r(s) - g_r(s-1) 
\right\}.
\label{expA14}
\end{split}
\end{equation}
Here, $ g_r (s)$ is a function of the form
\begin{equation}
 g_r (s) = \frac{c_{r,-2}}{s-i r \lambda - 2}
+ \frac{c_{r,-1}}{s-i r \lambda - 1}
+ \frac{c_{r,0}}{s-i r \lambda }
+ c_{r,1} s + c_{r,2} s^2 + c_{r,3} s^3,
\end{equation}
and $d_r$, $c_{r, -2}, \ldots, c_{r,3}$ are independent of~$s$. 
In order to integrate the term with~$d_r$, let us temporarily
consider integrating the modified function
\begin{equation}
F_p(s) =
 \frac{ e^{i \pi s}}{
\cos\{ \pi(\mu + s) \} \cos \{\pi (-\mu + s)\}
\sin \{\pi (s - i r \lambda )\}
}
 \frac{1  }{(s-i r \lambda )^p},
\label{p-modified}
\end{equation}
where the power~$p$ is a constant that satisfies $p > 1$. 
Closing the contour path of~$s$ on the left half-plane
with a semicircle of infinite radius 
(here consider an arc that does not pass through any of the poles,
and taking the infinite radius limit in a discontinuous manner),
one can check that the integral of~$F_p(s) $ 
along the arc vanishes due to $p > 1$.

Here, in addition to the requirements for the integration contour of~$s$
explained below (\ref{MellinBarnes}), and at (\ref{maru-ro}),
(\ref{maru-ha}),
we further demand the contour to pass to the left of 
$s = \frac{1}{2} \pm \mu, \, \frac{3}{2} \pm \mu $.
Such a path always exists for $\mu $ satisfying~(\ref{muconst}).  
Then the integral of~$F_p(s) $ can be obtained by summing up
its infinite series of residues at the simple poles:
\begin{equation}
 s = -\frac{1}{2} \pm \mu - n, \, \, i r \lambda - 1 - n,
\label{polesA17}
\end{equation}
where $n = 0, 1, 2, \dotsc$.
Then we take the limit $p \to 1$ of the integrated result
in order to obtain the integral of the $d_r$ term in (\ref{expA14}), 
which can be checked to take the form
\begin{equation}
\begin{split}
\lim_{p \to 1} \int^{i\infty}_{-i\infty} \frac{ds}{ 2\pi i} 
F_p(s)
= &- \frac{\gamma e^{-\pi r \lambda}}{\pi  \cos\{ \pi (\mu + ir 
 \lambda ) \}  \cos \{ \pi (\mu - i r \lambda ) \} }
\\
& 
- \frac{i}{\pi \sin (2 \pi \mu)}
\left[
\frac{e^{-i \pi \mu} \, \psi (\frac{1}{2} + \mu + i r \lambda)}{\cos
\left\{ \pi (\mu + i r \lambda) \right\} } 
- 
\frac{e^{i \pi \mu} \, \psi (\frac{1}{2} - \mu + i r \lambda)}{\cos
\left\{ \pi (\mu - i r \lambda) \right\} } 
\right].
\label{temrsfromnonXi1}
\end{split}
\end{equation}
Here, $\gamma$ is Euler's constant.

As for integrating~$g_r (s) - g_r (s-1)$ in (\ref{expA14}), 
we shift the variable for $g_r (s-1)$ by $s \to s+1$, giving
\begin{equation}
\left(
 \int^{i \infty}_{-i \infty} 
-  \int^{i \infty  - 1}_{ -i \infty -1} 
\right)
\frac{ds}{2 \pi i}
\frac{e^{i \pi s} g_r(s) }{
\cos\{ \pi(\mu + s) \} \cos \{\pi (-\mu + s)\} \sin \{\pi (s - i r \lambda )\}
}.
\label{temrsfromnonXi2}
\end{equation}
This can be evaluated by computing the residues of poles in the
region sandwiched by the original integration contour and the shifted
one (the choice of the contour was discussed above~(\ref{polesA17})),
which are simple poles at 
\begin{equation}
 s = i r \lambda - 1, \, \, -\frac{1}{2} \pm \mu .
\end{equation}

Summing up the contributions from (\ref{temrsfromXi}),
(\ref{temrsfromnonXi1}), (\ref{temrsfromnonXi2}), and integrating
over~$r$, we arrive at the final result:
\begin{equation}
\begin{split}
&  \int 
d^3 k\, 
\left(k_z + e A_z \right)
\frac{e^{i \kappa \pi}}{2 k}
\left| W_{\kappa, \mu} (z) \right|^2 
\\
& =
 \frac{\pi  }{\tau^3} \lim_{\xi \to \infty}
\Biggl[
\frac{2\lambda }{3}  \xi^2
+  \frac{\lambda }{3}  \ln (2 \xi )
-  \frac{25\lambda }{36}  - \frac{i \pi \lambda }{6}
+ \frac{\mu^2\lambda }{3}   + \frac{\lambda^3}{15} 
\\
& \quad
+\frac{45 + 4 \pi^2 (-2+3 \lambda^2 + 2 \mu^2)}{12 \pi^3}
\frac{\mu \cosh (2 \pi \lambda )}{\lambda \sin (2 \pi \mu)}
 -\frac{45 + 8 \pi^2 (-1+9 \lambda^2 +  \mu^2)}{24 \pi^4}
\frac{\mu \sinh (2 \pi \lambda )}{\lambda^2 \sin (2 \pi \mu)}
\\
& \quad
+  \int^1_{-1} dr\, 
\frac{i \lambda }{16 \sin (2 \pi \mu )}
\left\{
-1+4 \mu^2 + (7 + 12 \lambda^2 - 12 \mu^2) r^2 - 20 \lambda^2 r^4
\right\}
\\
& \qquad \qquad \quad
\times
\left\{
\left( e^{2 \pi r \lambda }+ e^{2 \pi i \mu} \right)
\psi \left( \tfrac{1}{2} + \mu + i r \lambda  \right)
-
\left( e^{2 \pi r \lambda }+ e^{-2 \pi i \mu} \right)
\psi \left( \tfrac{1}{2} - \mu + i r \lambda  \right)
\right\}
\Biggr].
\label{A.21}
\end{split}
\end{equation}
We see that the three-dimensional momentum integral for the current has
quadratic and logarithmic divergences. 
(The cutoff~$\xi$ is related to that in (\ref{rawJ}) by $\xi = - \tau  \zeta $.)
Let us also note that the integral for the values of~$\mu$ excluded
in~(\ref{muconst}) can be evaluated by taking the limits of the
result~(\ref{A.21}). 
It can further be checked that the imaginary term~$-\frac{i \pi \lambda}{6}$ in
the second line cancels with the imaginary part of the $r$-integral in
the forth and fifth lines, making it evident that the result is real.
Thus, with a normalization factor, we obtain the expression~(\ref{rawJ}).


\end{document}